\documentclass[traditabstract]{aa}
\usepackage{natbib}
\usepackage{graphicx}
\usepackage{txfonts}
\usepackage{lscape}

\begin{document}
\title{Properties of radio-loud quasars in Sloan Digital Sky Survey}
\authorrunning{Gaur et al.}
\titlerunning{Properties of radio-loud quasars}
\author{Haritma Gaur,$^{1,2}$ Minfeng Gu,$^{1}$ S. Ramya,$^{3}$ Hengxiao Guo,$^{4,5}$   }

   \institute{
Key Laboratory for Research in Galaxies and Cosmology, Shanghai Astronomical Observatory,
Chinese Academy of Sciences, 80 Nandan Road, Shanghai 200030, China (harry.gaur31@gmail.com)
\and
Aryabhatta Research Institute of Observational Sciences (ARIES), Manora Peak, Nainital -- 263002, India
\and
Indian Institute of Astrophysics (IIA), Bangalore, India
\and
Department of Astronomy, University of Illinois at Urbana-Champaign, Urbana, IL 61801, USA
\and
National Center for Supercomputing Applications, University of Illinois at Urbana-Champaign, 605 East Springfield Avenue, Champaign, IL 61820, USA
}

\abstract{We  present a study of a sample of 223 radio loud quasars (up to redshift $<$0.3) in order to investigate their spectral 
properties.  Twenty-six of these radio loud quasars are identified as Flat Spectrum Radio Quasars (FSRQs) and fifty-four are identified 
as Steep Spectrum Radio Quasars (SSRQs) based on their radio spectral index.
We study the [O III] line properties of these quasars to investigate the origin and  properties of blue wings 
(shift of the profile towards lower wavelengths) and blue outliers (shift of the whole spectroscopic feature).
 Most of the quasars show blue wings with velocities up to 420 km $s^{-1}$.
We find that around 17\% of the quasars show outliers with velocities spanning 419 to -315 km $s^{-1}$. 
Finally, we revisit the $\it M_{\rm BH} - \sigma$ relation of our sample using [S II]$\lambda$6716, 6731 and [O III] linewidths as
surrogates for stellar velocity dispersions, $\sigma$, to investigate their location on the $\it M_{\rm BH} - \sigma$ relation
for quiescent galaxies.
Due to strong blending of [S II] with $\rm H_{\alpha}$, we could estimate $\sigma_{[\rm SII]}$ of only 123 quasars.
 We find that the radio-loud quasars do not show a relationship
between $\it M_{\rm BH}$ and $\sigma_{\rm [SII]/[OIII]}$ up to a redshift of 0.3, although they cluster around the local relation.
We find an overall offset of 0.12$\pm$0.05 dex of our sample of radio loud quasars from the  $\it M_{\rm BH} - \sigma$ 
relation of quiescent galaxies. Quasars in our highest redshift bin (z=0.25-0.3) show a deviation of $\sim$0.33 $\pm$ 
0.06 dex with respect to the local relation. 
}

\keywords{galaxies: active -- galaxies:  jets -- quasars: emission lines}

\maketitle

\section{Introduction}

Quasars are classified as radio loud and radio quiet. The radio loudness parameter ($R$) is conventionally defined as the ratio of the
radio luminosity at 5 GHz to the optical luminosity at 4400 $\rm \AA$ (Kellermann et al. 1994). Then 10--15 \% of  quasars are called
radio loud, with $R \ge 10$. Radio loud quasars possess powerful radio jets that extend from sub-pc to well outside the 
galaxy and sometimes to Mpc scales. The remaining sources are radio 
quiet, with much weaker radio jets that are mostly confined within the host galaxy when they are detected at all (Kellermann et al. 2016; Padovani 2016).

Radio loud quasars are further divided into compact flat spectrum quasars ($\alpha > -0.5$) and extended steep spectrum
quasars ($\alpha < -0.5$) where the spectral index $\alpha$ is defined as $f_{\nu} \propto \nu^{-\alpha}$ with $f_{\nu}$ being the flux density at frequency
$\nu$. It has been found that a radio loud quasar consists of a compact central core and two extended lobes.
The spectrum of the central core is flat and spectrum of the lobes is steep. Hence, the radio loud quasar appears as a Flat 
Spectrum Radio Quasar (FSRQ) when it is core dominated and appears as an Steep Spectrum Radio Quasar (SSRQ) when it is lobe dominated.
The relativistic beaming model for radio sources unifies core dominated and lobe dominated sources by means of orientation (Blandford
\& Rees 1979) where core dominated objects are those viewed close to the jet axis, while lobe dominated objects are those viewed 
at larger angles.

Radio-loud quasars usually have powerful jets. Although the mechanism of jet formation still under debate, it has been proposed 
that the jet provides substantial feedback that could affect the circumnuclear environment, at the galaxy scale and even at larger, 
galaxy cluster, scales (McNamara \& Nulsen 2007). Recent observations show evidence for AGN (Active Galactic Nuclei) 
feedback in the form of massive large scale outflows (Fabian et al. 2012;  Greene et al. 2012; Scannapieco et al. 2012;
Zakamska \& Greene 2014). Also, the discovery of many massive molecular outflows has given support to AGN feedback models
(Feruglio et al. 2010; Aalto et al. 2012; Maiolino et al. 2012) and found relations between outflow rates and various AGN
properties. Cicone et al. (2014) found that outflow rates correlate with the AGN power.
The [O III] emission line shows asymmetry and blueshifts, indicating that NLR is undergoing an organized
outflow (Zakamska \& Greene 2014; Xu \& Komossa 2009).  The [OIII] line profiles are generally characterized by two distinct components.
The first component represents the line core, having almost same redshift as that of the host galaxy, while the second component
is systematically blueshifted (sometimes redshifted) and has a higher FWHM than the core component. The second component is usually called a Blue Wing (BW)
and is associated with a gas outflow in the narrow line region (NLR) (Komossa et al. 2008). Blue wings are thought to be generated 
by strong winds in the sources having high Eddington ratios (Proga et al. 2000). Sometimes, the first component also shows
blueshift with respect to their rest frame wavelength (Zamanov et al. 2002) and such sources are called blue outliers (BOs)
(Komossa et al. 2008). 

The formation of these BOs are not well understood in AGNs but in radio loud quasars, powerful relativistic
jets could possibly interact with the NLR and lead to such BOs. In previous studies, it has been found that larger widths of 
narrow lines are found in AGNs which have powerful radio jets associated with them (Peterson 1997). The fast relativistic jet 
could accelerate the gas and release part of its energy into thermal energy of the surrounding gas (Pedlar et al. 1995). 
However, it has been found from simulations (Wagner \& Bicknell 2011; Wagner et al. 2012) that only powerful jets 
can affect the gas kinematics in the NLR and hence lead to the generation of BOs.

It is well known that the galaxies having massive bulges contain central Super Massive Black Holes (SMBHs).
The studies of the correlation between the mass of the SMBH,  $M_{\rm BH}$, and the host stellar velocity dispersion, $\sigma$, which is called the 
$M_{\rm BH} - \sigma$ relation, is of fundamental importance to understand the galaxy formation and evolution. Earlier 
studies found a close connection between the black hole mass and bulge stellar velocity 
dispersion  (Ferrarese \& Merritt 2000; Gebhardt et al. 2000; Tremaine et al. 2002; Lauer et al. 2007; Kormendy \& 
Ho 2013; Shen et al. 2015 and references therein) and a close link between the black hole mass and bulge formation and
growth (Marconi \& Hunt 2003; Haring \& Rix 2004; Haehnelt \& Kauffmann 2000). 

An updated $\it M_{\rm BH}-\sigma$ relation for the local inactive galaxies is given by Kormendy \& Ho (2013) for classical
bulges/elliptical galaxies as:

\begin{equation}
\begin{array}{l}
\frac{M_{\rm BH}}{10^{9} M_{\odot}} = \Big(0.310 ^{+0.037} _{-0.033}\Big)  \Big(\frac {\sigma} {200~\rm km s^{-1}}\Big)^{4.38\pm0.29}.
 \end{array}
\end{equation}

The location of Active Galactic Nuclei (AGNs) 
on the $\it M_{\rm BH}-\sigma$ plane of quiescent galaxies is of great importance as it would provide strong constraints 
on their evolution. Various classes of AGNs, such as narrow-line Seyfert 1 galaxies, radio-quiet, radio-loud quasars, 
and intermediate SMBH, sometimes were found to be
on the relation and sometimes to be off it (Nelson 2000; Boroson 2003; Shields et al. 2003; Grupe \& Mathur 2004; Bonning
et al. 2005;  Salviander et al. 2007; Komossa \& Xu 2007; Shen et al. 2008; Gu et al. 2009; Ramya et al. 2011; Woo et al. 2013;
Salviander \& Shields 2013; Bennert et al. 2014; Subramanian et al. 2016). 

To firmly establish the $M_{\rm BH}-\sigma$ relation for different samples, accurate and uniform estimations of both $M_{\rm BH}$ 
and $\sigma$ are required; however, this is often difficult to achieve. 
$M_{\rm BH}$ is commonly estimated by combining 
the measured line width of broad emission lines like $\rm H\alpha$, $\rm H\beta$, Mg II, C IV with the reverberation mapped based empirical relationship 
between the broad line region (BLR) radius and continuum and/or line luminosity (Kaspi et al. 2000; Greene \& Ho 2005; Peterson 
et al. 2004). In AGNs, the nuclei is usually much brighter than the host galaxy itself. Therefore, the fundamental 
limitation is that the stellar velocity dispersions can not be directly measured, except in a few low luminosity AGNs (Greene \& Ho 2005; Shen et al. 
2015 and references therein).  Instead of directly measuring $\sigma$, the line width of narrow  [O III]
$\lambda$5007 is commonly used as a surrogate for $\sigma$ (Komossa et al. 2007; Gu et al. 2009; Salviander \& Shields 2013 ).
There is a significant correlation between NLR gas and the gravitational potential 
of bulge of the host galaxy (Nelson 2000), however the $\sigma_{[OIII]}$
shows more scatter than the stellar velocity $\sigma$ on the Faber-Jackson relation (Nelson \& Whittle 1996; Bonning et al. 2005; Xiao et al. 2011). 

From previous studies, it has been found that the radio-loud quasars deviate from the $M_{\rm BH}-\sigma$ relation for quiescent
galaxies (Bian \& Zhao 2004; Bonning et al. 2005; Bian et al. 2008; Shen et al. 2008; Gu et al. 2009).
Radio loud quasars mostly settle above the relation, i.e., have larger black hole masses for a given stellar velocity dispersion. 
In all of this previous work, the[O III] line width is used as surrogate for stellar velocity dispersion, hence can include quasars 
up to 0.8 redshift. As we know that [O III] line often suffers from asymmetry in the line profile and a strong blue-shifted 
wing component due to outflows (e.g. Boroson 2005; Bae \& Woo 2014), the uncertainty of this proxy can be very 
large, as shown by the direct comparison between [O III] line width and the measured stellar velocity dispersion (Woo et al. 2006; 
Xiao et al. 2011). This illustrates that in order to use the [O III] line width as a proxy for $\sigma$ in active galaxies, 
the blue wing should be properly removed. It has also been proposed that because of the complexity and asymmetry of this line, other 
low ionization emission lines like [S II] $\lambda$6716, 6731, [N II] $\lambda$6584,6548 and [O I] $\lambda$6300 from the NLR can be used as 
proxies for $\sigma$ (Nelson \& Whittle 1996; Greene \& Ho 2005).
Because the [N II] line is usually blended with strong $\rm H\alpha$ (e.g., Zhou et al. 2006), and [O I] is usually weak, the [S II] line 
was proposed to be a better indicator of $\sigma$ (Greene \& Ho 2005). 

In this work, we study a sample of radio loud quasars to investigate their spectral properties to address their distributions of black hole mass,
Eddington ratios, radio luminosity, etc. By calculating the radio spectral index, we classify the sample of radio loud quasars into FSRQs and 
SSRQs. We also investigate the spectral properties of [O III] lines of our sample to search if BOs are present in radio loud quasars, and 
if present, whether relativistic jets are responsible for their production.
Finally, we revisit the $\it M_{\rm BH}-\sigma$ relation for radio-loud quasars using [S II] and [O III] line widths as
surrogates of $\sigma$, to investigate their location on the $M_{\rm BH}-\sigma$ relation for quiescent galaxies. Also, 
we try to study whether there is difference in the location of FSRQs and SSRQs on the $M_{\rm BH}-\sigma$ relation;
The cosmological parameters, $H_{0}=70~ \rm km~ s^{-1} Mpc^{-1}$, $\Omega_{m}$=0.3, $\Omega_{\lambda}$ =0.7 are used throughout the
paper. The sample selection and data reduction are described in Sections 2 and 3, respectively. The results are 
presented in Section 4, while a discussion and our conclusions are given in Section 5.
\section{Sample Selection}

We select our sample of radio-loud quasars from the Sloan Digital Sky Survey (SDSS) Data Release (DR10) quasar catalogue, 
which consists of 105,783 bonafide quasars brighter than $M_{\rm i}=-22.0$ and has at least one broad emission line width 
larger than 1000 $\rm km ~s^{-1}$ (Schneider et al. 2010). 
SDSS optical spectra cover the wavelength range $3800-9200~ \rm \AA$ with a spectral resolution of  $\sim1850-2200$ 
(i.e. Schneider et al. 2010).  
The radio flux densities at 1.4-GHz are tabulated from the Faint Images of the Radio Sky (FIRST) 
radio catalogue (Becker, White \& Helfand 1995) and the radio-loudness parameter $R=f_{\rm 6~cm}/f_{2500}$ then estimated, where $f_{\rm 6~cm}$ 
and $f_{2500}$ are the flux densities at 6 cm and 2500 $\rm \AA$ at the rest frame of the source, respectively (see details in Shen et al. 2011). 

We select radio loud quasars with redshift z $\leq$ 0.3 based on the criteria that $R\ge10$ as given in Shen et al. (2011).
In estimating $R$, we compute the 5 GHz flux density
from the FIRST integrated flux density assuming a power law slope of 0.5 as done in Shen et al. (2011).
This radio loudness calculation can be questioned as the effect of relativistic beaming affects both radio and optical emission, 
especially for FSRQs, and can do so differentially. However, an accurate correction for an individual object is rather difficult.

 Due to flux limit of FIRST (1 mJy), some radio-loud quasars may not be
found through cross matching, causing biases against high-z sources. In order to check this bias, we plot percentage of sources
detected by FIRST i.e. (Sources detected with FIRST/Total sources observed in that redshift range) along the redshift,
to see whether the fraction is decreasing with increasing redshift. It can be seen from Figure 1. that there
is no correlation between percentage of sources detected with FIRST versus redshift (upto z=0.3).
Hence, it is difficult to conclude that the sample is biased against more distant objects. 
Also, in order to check whether the sources which are non detected by FIRST are indeed at R$<$10, we calculated
upper limits of radio loudness (R) of non-detected sources by measuring $F_{5GHz}$ using the limiting flux of FIRST (i.e. 1mJy).
We found that all the sources have radio loudness less than 10. \\

We know that [S II] lines are lower ionization lines and need a high signal-to-noise ratio to decompose them. Therefore, we extract
only those quasars which have mean $\rm S/N>$10. These two steps resulted in a sample containing 230 quasars. 
In many spectra, the [S II] lines are strongly blended with $\rm H\alpha$ emission line, hence we could estimate $\sigma_{[SII]}$ of only 123
quasars. Since most broader $\rm H\alpha$ profiles are associated with more massive quasars, which could lead to a selection bias
in the way that we probably excluded some of the most massive galaxies. In radio loud quasars, broader $\rm H\alpha$ profiles affect the
estimation of [S II] line widths. Hence, we also estimate [OIII] line widths for the whole sample.
We exclude 7 spectra upon visual inspection due to spectral defects in the region of [OIII], hence we are left with a total of 223 spectra.
Only 80 quasars out of these 223 quasars 
have well measured 5 GHz flux densities that allow us to determine uniform radio spectral indices; for them we obtained the 5 
GHz fluxes by cross-correlating with the Green Bank 6cm (GB6) catalogue (Gregory et al. 1996).  From this useful subsample, 26 quasars 
are identified as flat-spectrum radio quasars (FSRQs) 
with $\alpha_{\rm 1.4-5 GHz} \le 0.5$, and 54 objects as steep-spectrum radio quasars (SSRQs) with  $\alpha_{\rm 1.4-5 GHz} > 0.5$. 

\begin{figure}
\centering
\includegraphics[width=9cm , angle=0]{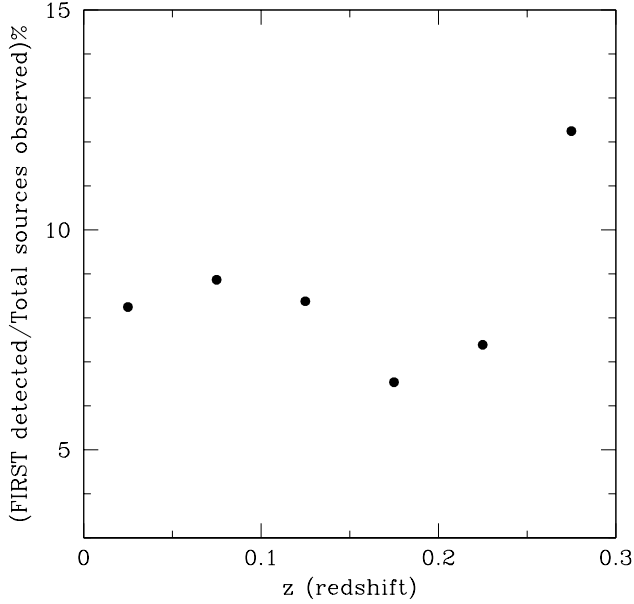}
\caption{Percentage of sources detected with FIRST versus redshift.}
\end{figure}

\begin{figure}
\centering
\includegraphics[width=9cm , angle=0]{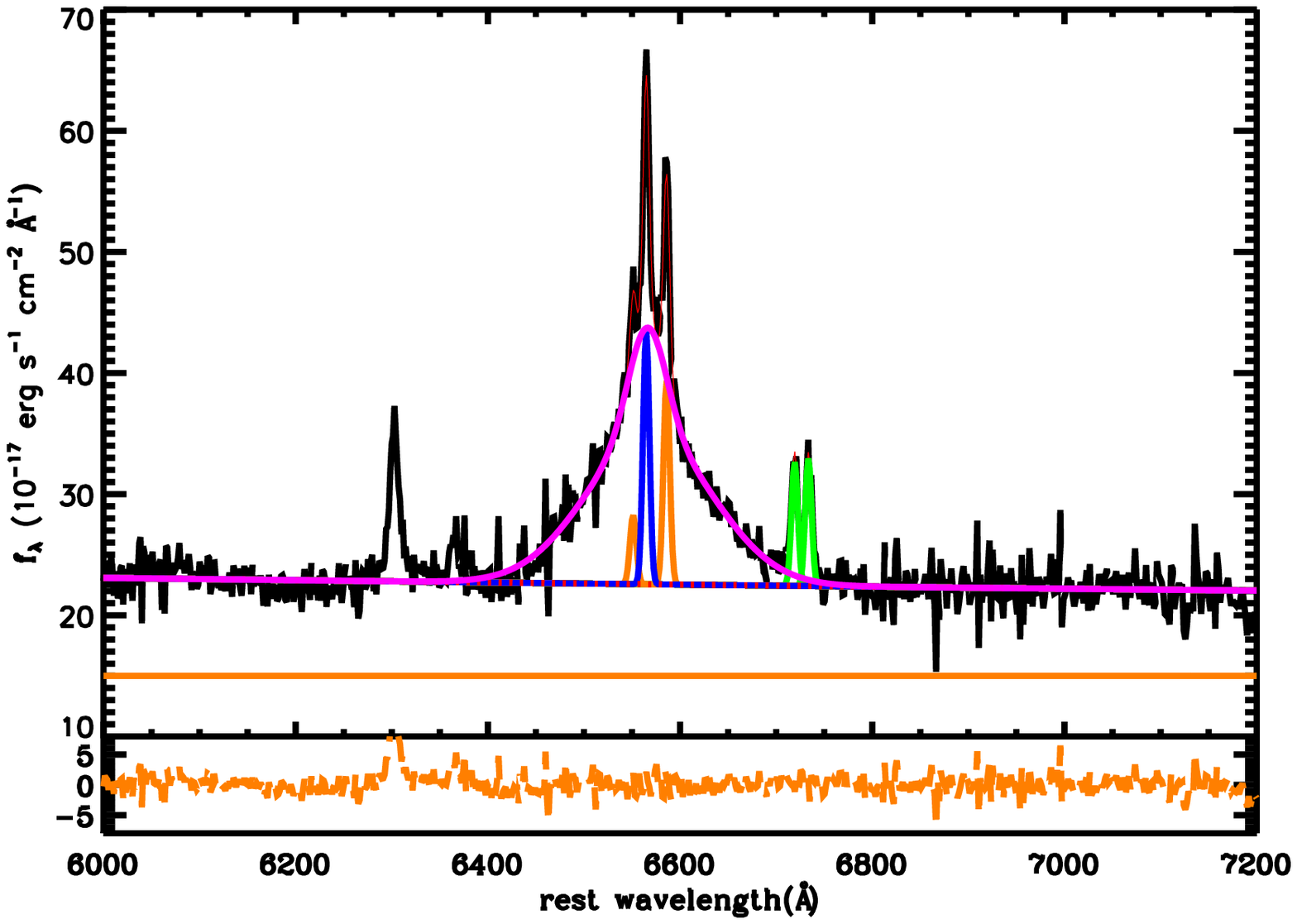}
\includegraphics[width=9cm , angle=0]{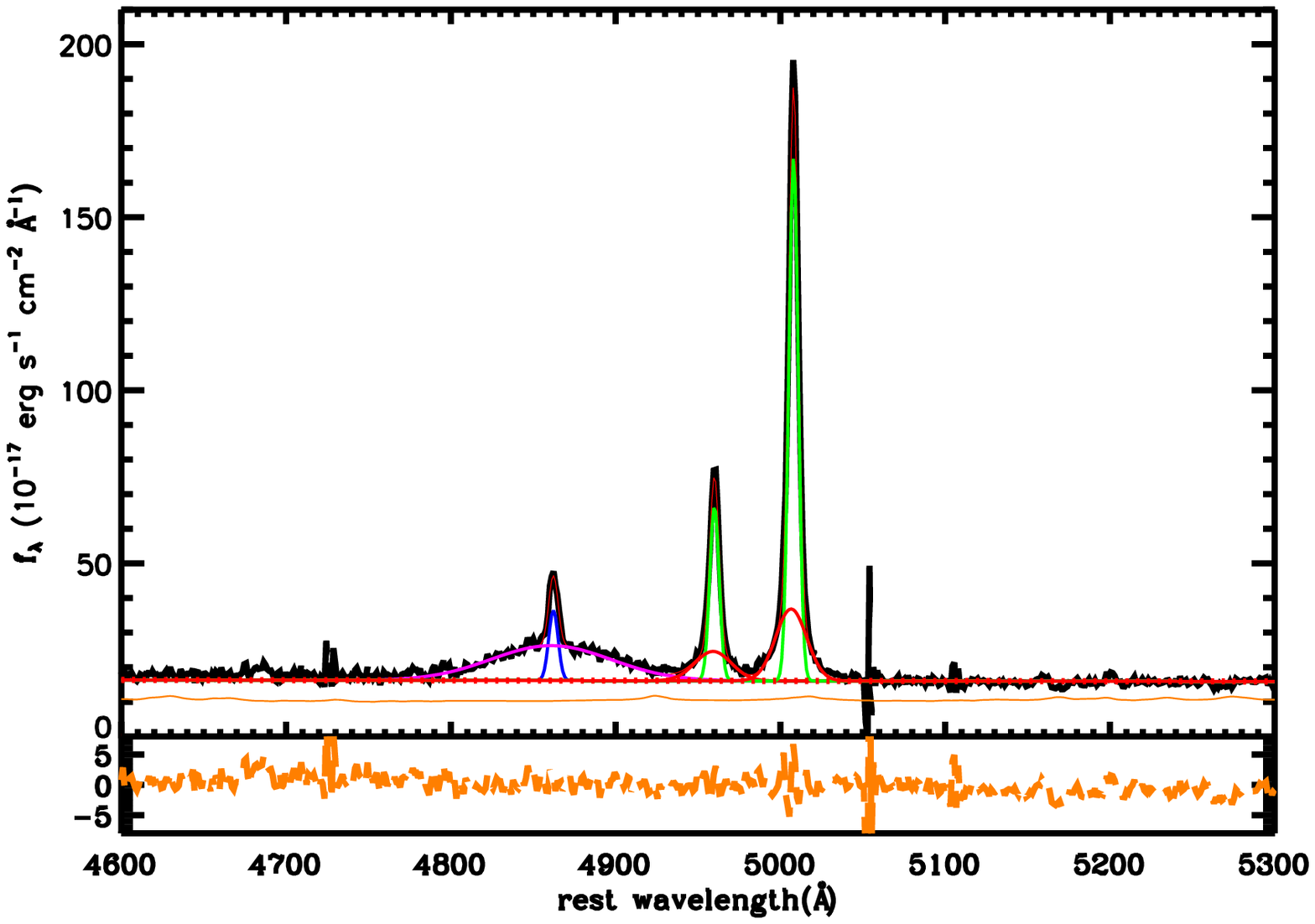}
\caption{Examples of fits to the spectra in $\rm H\alpha$ (upper panel) and $\rm H\beta$ (lower panel) regions, respectively. 
In both the panels, black lines indicate the observed spectrum, thick red lines indicate the complete fit to the spectra 
including all the Gaussian components. The power law continuum and the Fe II template are shown by red and orange lines, respectively. 
In the upper panel, broad and narrow $\rm H\alpha$ are shown by magenta and blue lines, respectively. [S II] and [N II] lines 
are shown by green and orange lines, respectively. In the lower panel, broad and narrow $\rm H\beta$ are shown by magenta and 
blue lines, respectively. Narrow core of [O III] is shown by green line
and the wing component is represented by red line. The residuals of the fits are shown in the lower 
panels of both figures.}
\end{figure}

\section{Spectral Analysis}

The SDSS spectra are corrected for Galactic extinction using the reddening map of Schlegel, Finkbeiner \& Davis (1998). 
These spectra are then shifted to their rest wavelength by 
adopting redshift from header of the SDSS spectra. In order to fit the power-law continuum,
 we choose the wavelength range that is not affected by prominent emission lines. We use the
optical Fe II template from Veron-Cetty et al. (2004) which covers the wavelength range of $3535-7534~ \rm \AA$. The 
continuum and Fe II components are fit on the line-free spectral windows by minimizing $\chi^{2}$, and are then 
subtracted from the spectra (see details in Chen et al. 2009). In our sample of radio loud quasars, the nucleus is relatively 
bright with respect to the host galaxy, hence we are ignoring the host galaxy contribution to the spectrum. 

In order to estimate flux and velocity dispersion of [S II] and [O III] from the continuum subtracted spectra,
we fit local regions around $\rm H\alpha$  and $\rm H\beta$ following the procedure of Shen et al. (2011).
For $\rm H\alpha$, we fit the wavelength range of $6000-7180 ~\rm \AA$. The narrow components of $\rm H\alpha$, 
[N II] $\lambda$$\lambda$ 6548, 6584, [S II] $\lambda$$\lambda$ 6717, 6713 are modeled with a single Gaussian profile and 
their line widths were tied to the same.
The flux ratio of [N II] doublets are fixed to be 2.96 (Osterbrock et al. 1989). Following Hao et al. (2005), we impose an 
upper limit on the line width of narrow components,
 $\rm FWHM< 1200~ km~ s^{-1}$. The broad $\rm H\alpha$ component is modeled with multi-Gaussian components, i.e.,
 starting with a single Gaussian with $\rm FWHM> 1200~ km~ s^{-1}$ up to three multiple Gaussians, each with $\rm FWHM> 1200~ km~ s^{-1}$.
New Gaussian components are added one at a time if they lead to a reduction of $>20 \%$ in $\chi^{2}$ following Xiao et al. (2011).
 Generally, one or two Gaussians are sufficient to fit the broad $\rm H\alpha$ profile. However, quasars with 
asymmetric, double peaked profiles, or very broad wings, require more than two Gaussian components.    

Similar fitting is performed for $\rm H\beta$ and [O III] lines in the wavelength range $4200-5300 ~\rm \AA$. Each of the [O III] $\lambda$$\lambda$ 
4959,5007 line is model by a double Gaussian model, i.e., a core (with higher flux) and a mostly blue shifted wing component 
(with lower flux) (e.g. Heckmann et al. 1981; Greene \& Ho 2005; Komossa et al. 2008). The FWHM of narrow $\rm H\beta$ line is
tied to that of the [O III] core, with an upper limit of $1200 \rm~km ~s^{-1}$. The flux ratio of [O III] $\lambda$4959 to [O III] $\lambda$5007 is constrained to be 1:3 (Osterbrock 1989). 
 As for $\rm H\alpha$, the broad $\rm H\beta$ component is model with multi-Gaussian components (up to three), each with 
FWHM $>$ 1200 km $s^{-1}$.  In adding new Gaussian component, we followed the method adopted in Greene \& Ho (2005), 
Section 3.2. Also, we manually check the fiting results of some spectra using F-test while adding new Gaussian components and
 found that F-test values are significant when new $\chi^{2}$ lead to a reduction of $>20\%$. Hence, we adopted this 
more stringent criterian to make the fitting model more simple 
and a new Gaussian profile is added only if $\chi^{2}$ is reduced by $>20 \%$ (Xiao et al. 2011). Another 
method to fit the asymmetric profiles of $\rm H\beta$ component is to use a
Gauss Hermite series (van der Marel \& Franx 1993). The broad component of $\rm H\beta$ line are fit using a sixth order Gauss Hermite series.
More details are given in Park et al. (2012). We fit a few quasars in our sample using this method and find that both methods 
yield very similar results (Shen et al. 2011).

 Asymmetric profiles are clearly seen in [O III] but not detected
in [S II] profiles. We use {\it MPFITFUN} of {\it IDL} to fit the emission lines employing {\it GUASS1} program. The parameters
need to be fit for each Gaussian are centroid, peak value and sigma. 
To estimate uncertainties in the measured quantities from Single Epoch spectra, we generate 100 mock spectra by adding Gaussian noise
to the original spectrum using the flux density errors. Then, we fit these simulated spectra using the fitting procedure described above.
We estimate the standard deviation of the distribution of measurements from these 100 simulated spectra as the measurement uncertainty.
Uncertainties on fluxes are estimated by adding errors of peak and sigma in quadrature. 
Example of emission line profile fits in the $\rm H\alpha$ and $\rm H\beta$ regions are shown
in upper and lower panels of Fig 2, respectively.

\section{Results }

\subsection{Black hole mass estimation}

Black hole masses are derived using different methods in the literature (i.e. Czerny \& Nicolajuk 2010; Shen 2013, for
 recent reviews). For quiescent galaxies, $M_{BH}$ are estimated through simulations of galaxy stellar dynamics
(e.g., Gebhardt et al. 2000). For AGN, the reverberation mapping method is most accurate for measuring $M_{BH}$ (Blandford \& McKee 1982).
This method assumes that the BLR (Broad emission Line Region) is virialized and the motion of the emitting
clouds is dominated by the gravitational field of the Supermassive BH (e.g. Ho 1999; Wandel et al. 1999), i.e.

\begin{eqnarray}
 M_{\rm BH} = f \times \frac{R_{\rm BLR}V^2_{\rm BLR}}{G},
\end{eqnarray}
where $G$ is the gravitational constant, $R_{\rm BLR}$ is the radius of the BLR, $V_{\rm BLR}$ is the rotational velocity of the ionized gas
and $f$ is a dimensionless factor that accounts for the unknown geometry and orientation of the BLR.
When the continuum flux, which arises from 
the accretion disc or very close to it, varies with time this is later echoed by changes in flux of the
BLR, assuming that the BLR is powered by photoionization from the central source. 
Therefore, $R_{\rm BLR}$ is obtained by cross-correlation of the light curves which provides the time delay
between the continuum variations and the BLR variations. $V_{\rm BLR}$ is estimated from the width of the Doppler
broadened emission lines. This reverberation mapping technique requires high quality spectrophotometric monitoring
of AGNs over an extended period of time.  Values of $M_{BH}$ for over 50 AGNs have been estimated using this method (Kaspi et al. 2000;
Peterson et al. 2004; Bentz et al. 2009). The uncertainty in the  $M_{BH}$ calculation through reverberation mapping method
is between 0.4--0.5 dex (Shen 2013).
 
The single-epoch virial methods 
assume that the BLR gas is virialized and follows 
the radius--luminosity relation of the form $R_{BLR} \propto L^{\alpha}$.
The coefficients of this relation are determined from estimates of a sample of AGNs for which reverberation mapping 
data is available. $V_{\rm BLR}$ in this method is estimated from the FHWM of broad $\rm H\alpha$ or $\rm H\beta$ 
emission lines. $R_{BLR}$ is estimated using the monochromatic continuum luminosity of the host galaxy at 5100 $\rm \AA$.
As the continuum luminosity is correlated with $L_{\rm H\alpha}$ and $L_{\rm H\beta}$  (Greene \& Ho 2005), the mass of the black
hole can be estimated using FWHM and luminosities of either of the Balmer lines. 
Greene \& Ho (2005) provide equations for the SMBH masses in terms of the $\rm H\alpha$ and $\rm H\beta$ lines: 
\begin{eqnarray}
\label{eqhb}
M_{\rm BH} =   (3.6\pm0.2)\times 10^6 \times \Big(\frac{L_{\rm H\beta}}{10^{42}\,{\rm erg\,s^{-1}}}\Big)^{(0.56\pm0.02)} \nonumber \\ \times  \Big(\frac{\rm FWHM_{H\beta}}{\rm 10^3\,km\,s^{-1}}\Big)^2 \hspace{2.8cm} [M_{\odot}],
\end{eqnarray}
and
\begin{eqnarray}
\label{eqha}
M_{\rm BH} =   2.0^{+0.4}_{-0.3}\times 10^6 \times \Big(\frac{L_{\rm H\alpha}}{10^{42}\,{\rm erg\,s^{-1}}}\Big)^{(0.55\pm0.02)} \nonumber \\  \times \Big(\frac{\rm FWHM_{H\alpha}}{\rm 10^3\,km\,s^{-1}}\Big)^{(2.06\pm0.06)} \hspace{1.25cm}  [M_{\odot}].
\end{eqnarray}

The FWHM of the combined profile of broad $\rm H\alpha$ is usually used to calculate the black hole mass 
because it normally has higher S/N compared to $\rm H\beta$.
The empirical relation between the radius of the BLR and the continuum
luminosity at 5100 $\rm \AA$ is also generally used to estimate the black hole mass (e.g., Kaspi et al. 2000).  
In radio-loud quasars, the optical continuum could be subjected to contamination by synchrotron radiation from relativistic jets that could 
 be partly beamed, particularly in FSRQs. Under such circumstances, the optical continuum could be boosted due to the non-thermal 
jet emission which could overestimate the true thermal component, in turn systematically overestimating the black hole masses 
(Kaspi et al. 2000; Greene \& Ho 2005; Chen et al. 2009).
This is illustrated in the relationship between the $5100~\rm \AA$ continuum luminosity $L_{5100}$ and broad $\rm H\alpha$ luminosity 
$L_{\rm H\alpha}$ in Fig. 3. The solid line represents relation of $L_{H\alpha}$--$L_{5100}$ given by Greene \& Ho (2005) derived for 
radio-quiet AGNs. While many quasars follow the relation of radio-quiet AGNs, a significant number of
quasars lie below the line with a larger $\it L_{\rm 5100}$ at fixed $L_{\rm H\alpha}$ than radio-quiet AGNs. This is likely due to  
contamination of non-thermal jet emission continuum luminosity in $L_{\rm 5100}$.

Besides the contamination of the jet emission to the continuum emission, the estimation of black hole masses may also be affected by BLR geometry.
It has been argued that the BLR in radio-loud AGNs can have a more disc-like geometry (Wills \& Browne 1986;
Vestergaard, Wilkes \& Barthel 2000). In this scenario, for smaller jet viewing angles, narrower broad lines will be observed, and thus
the black hole masses will be underestimated. This should be a particular issue for FSRQs, where the jet is moving toward us 
at a fairly small viewing angle (Jackson \& Wall 1999; Lacy et al. 2001; McLure \& Dunlop 2002).
Normally, it is difficult to correct the BLR inclination for each source, as we normally do not have information of their orientation
(Shen \& Ho 2014).

 Therefore, in our sample of radio-loud quasars, we estimate $M_{\rm BH}$ using the empirical relation
which utilizes the FWHM and luminosity of $\rm H\alpha$. A newer empirical relation to calculate the black hole mass
using the line width and luminosity of broad $\rm H\alpha$ line  is given by Reines, 
Greene \& Geha (2013). They updated equation (4) of Greene \& Ho (2005) using  the modified radio luminosity relationship 
of Bentz et al. (2013). The modified empirical relation is as follows:

\begin{equation}
\begin{array}{l}

\rm log~ \big(\frac{{\it M}_{\rm BH}}{{M}_{\odot}}\big) = log~ \epsilon + 6.57 + 0.47~ log~ \Big(\frac{{\it L}_{H\alpha}}{10^{42}\,erg\,s^{-1}}\Big)\\
+2.06 ~\rm log ~\Big(\frac{{\rm FWHM}_{H\alpha}}{10^3\,km\,s^{-1}}\Big)

\end{array}
\end{equation}
where $\epsilon$ is the scale factor that depends on BLR geometry and spans a range of $\sim0.75-1.4$
(e.g. Onken et al. 2004; Grier et al. 2013). Here, we assumed $\epsilon$ =1.

 Shen (2013) estimated the uncertainty in the calculation of black hole mass using the
single-epoch (SE) virial BH mass estimators and found that the dominant uncertainty in log $M_{\rm BH}$
is the systematic uncertainty which can be $\sim$0.5 dex.

 The BLR luminosity $L_{BLR}$ is derived following Celotti, Padovani \& Ghisellini (1997)
by scaling the strong broad emission lines $H_{\beta}$ to the quasar tempelate spectrum of
Francis et al. (1991), in which $Ly_{\alpha}$ is used as a flux reference of 100.  From the BLR
luminosity, we estimated the disc bolometric luminosity as $L_{Bol}$=30 $L_{BLR}$ (Xu, Cao \& Wu 2009).
The bolometric luminosity of our sample of quasars varies between log($L_{bol}$)=44.69--46.38 ($\pm$0.005) (erg s$^{-1}$).
Black hole masses, radio loudness and Eddington ratio distributions of the whole sample of quasars, as well as those identified 
as FSRQs and SSRQs are shown in the top and bottom panels of Fig 4. 

 Black hole masses range between $10^{6}$ to $ 10^{9.5}$ $M_{\odot}$ for the whole sample. The black hole masses of FSRQs range between 
$10^{7.38}$ to $10^{9.40}$ $M_{\odot}$ while the range is $10^{6.37}$ to $10^{9.45}$ $M_{\odot}$ for SSRQs. 
 However, as mentioned above, due to the presence of systematic uncertainty of $\sim$0.5 dex in $M_{\rm BH}$ estimation,
there is likely no significant difference between the $M_{\rm BH}$ ranges quoted for flat-spectrum and steep-spectrum quasars. 
The logarithms of the radio loudness parameter of FSRQs range between 1.044 to 3.814 and of SSRQs between 1.011 to 4.340. The Eddington ratio log($L_{Bol}/L_{Edd}$) 
values for FSRQs are between -2.292 to -0.251 and while SSRQs are found between -2.411 to -0.519.
The table in the appendix lists all the values and estimates from the optical measurements of our radio-loud sample.

\begin{figure}
\centering
\includegraphics[width=7.5cm , angle=0]{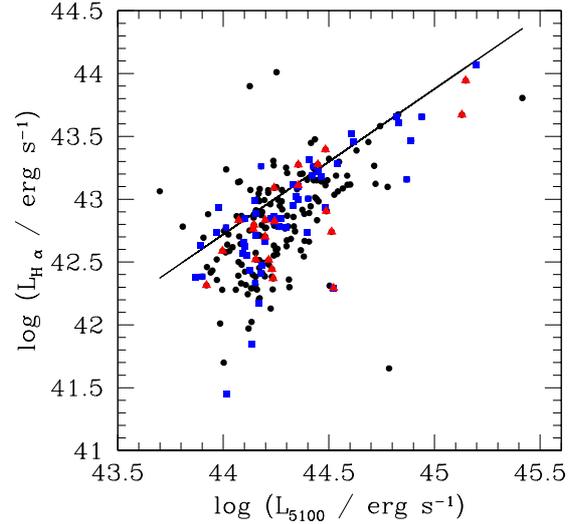}
\caption{Correlation between $\it L_{\rm H\alpha}$ w.r.t. $\it L_{\rm 5100}$ of radio loud sample. The solid line represents the relation
of $L_{H\alpha}$--$L_{5100}$ given by Greene \& Ho (2005) derived for radio-quiet AGNs. Unclassified radio loud quasars, 
FSRQs and SSRQs from our AGN sample are represented by black, red and blue symbols, respectively.
}
\end{figure}

\begin{figure*}
\centering
\includegraphics[width=5cm, angle=0]{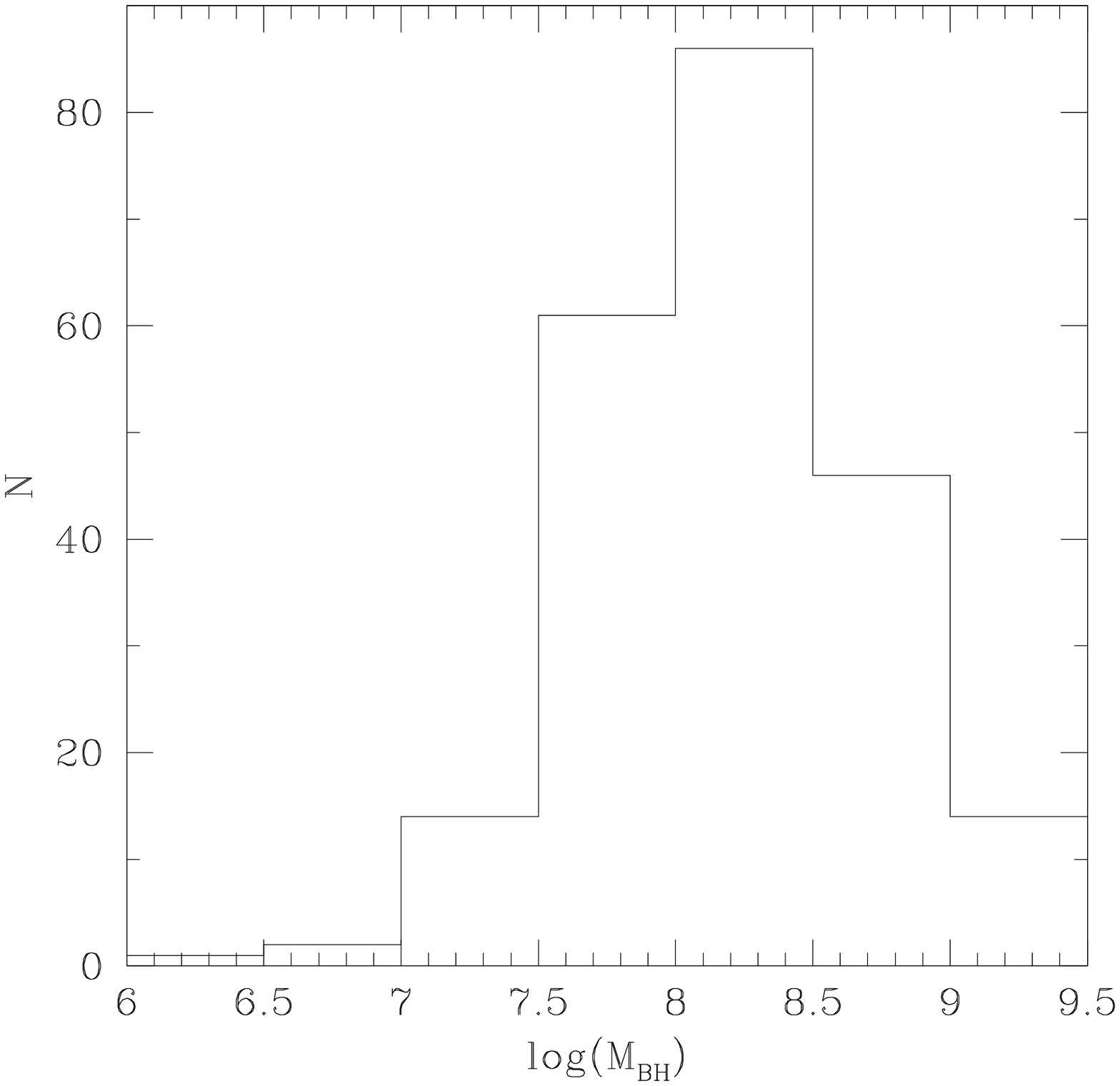}
\includegraphics[width=5cm, angle=0]{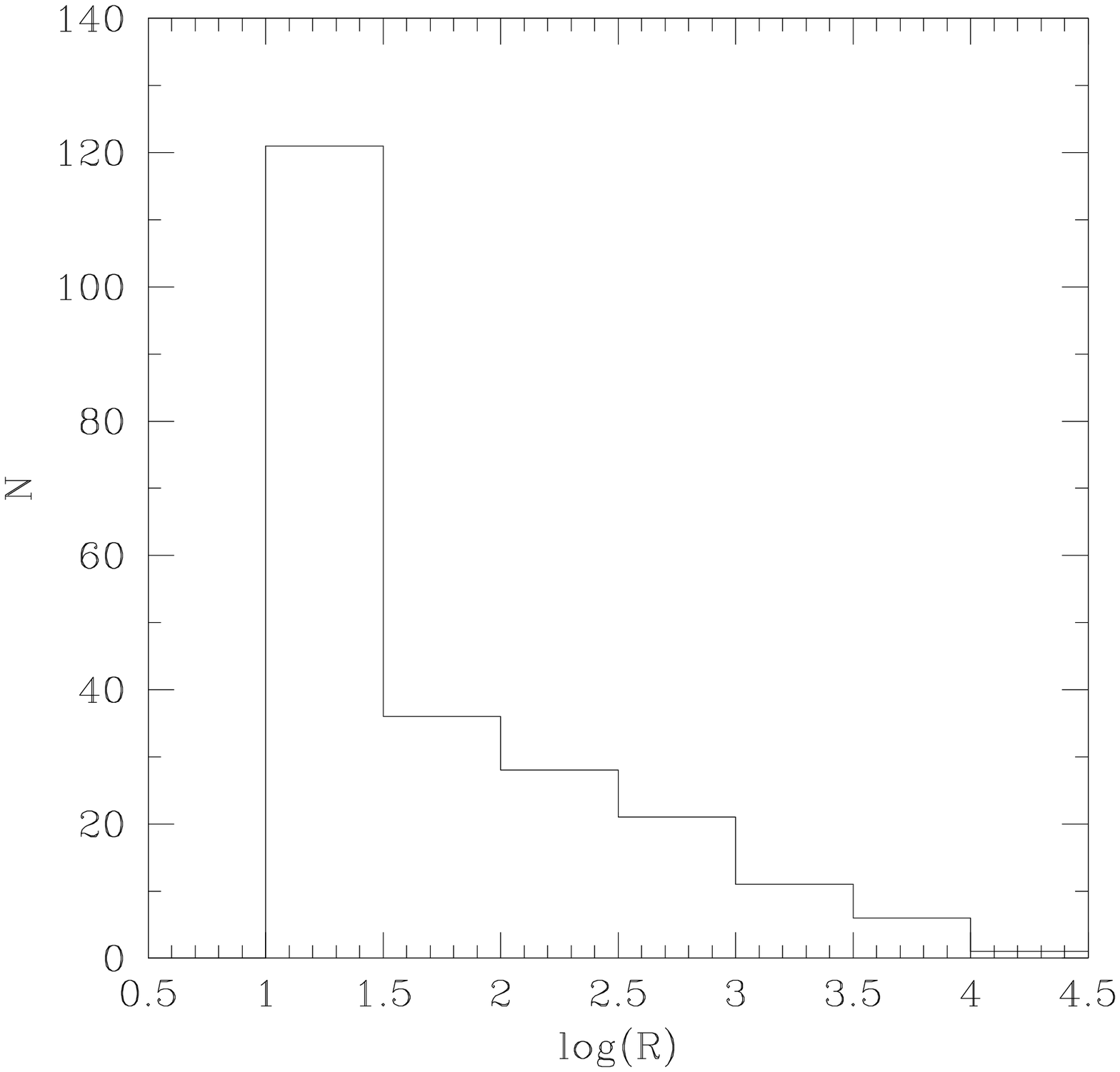}
\includegraphics[width=5cm, angle=0]{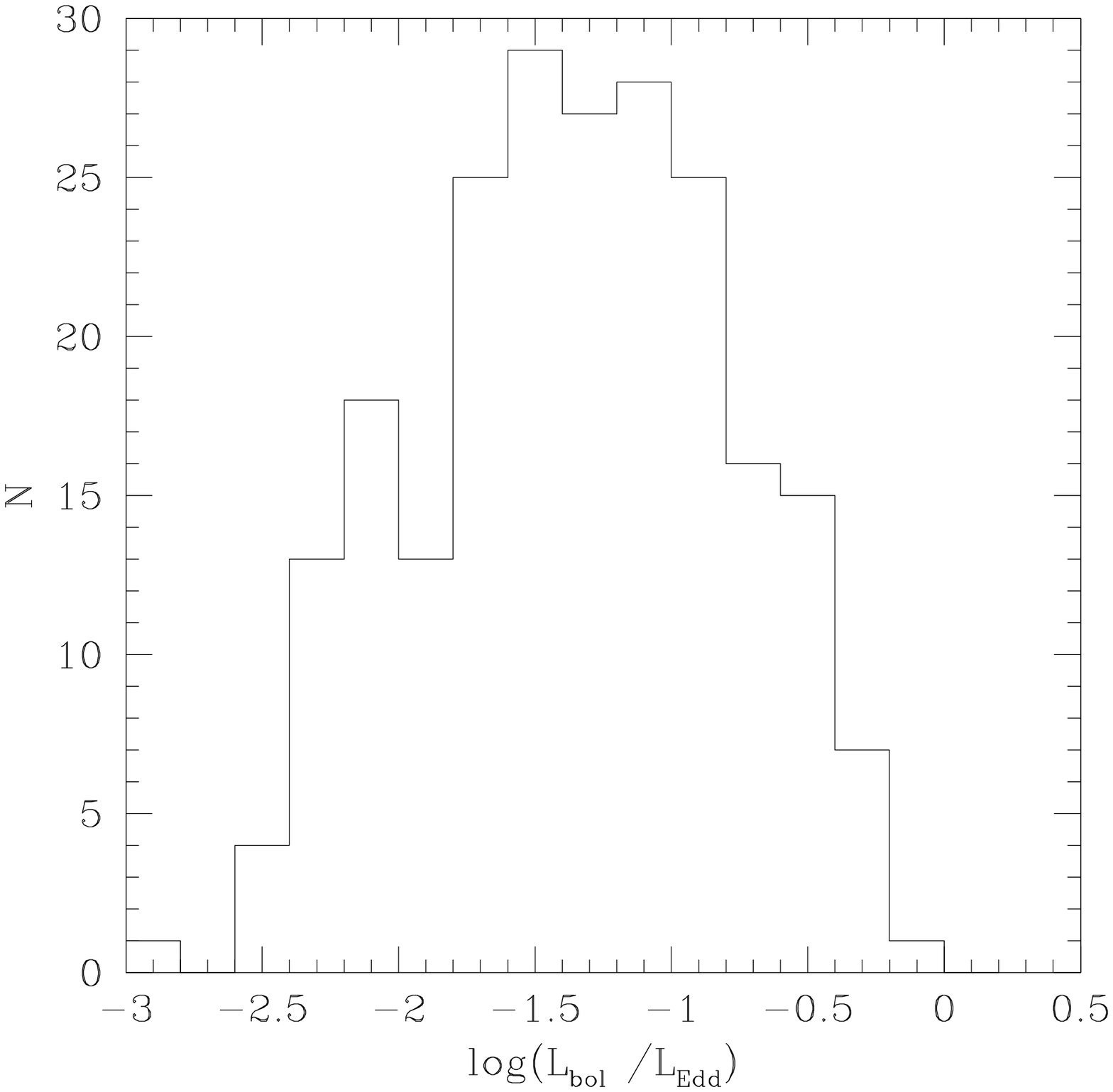}
\includegraphics[width=5cm, angle=0]{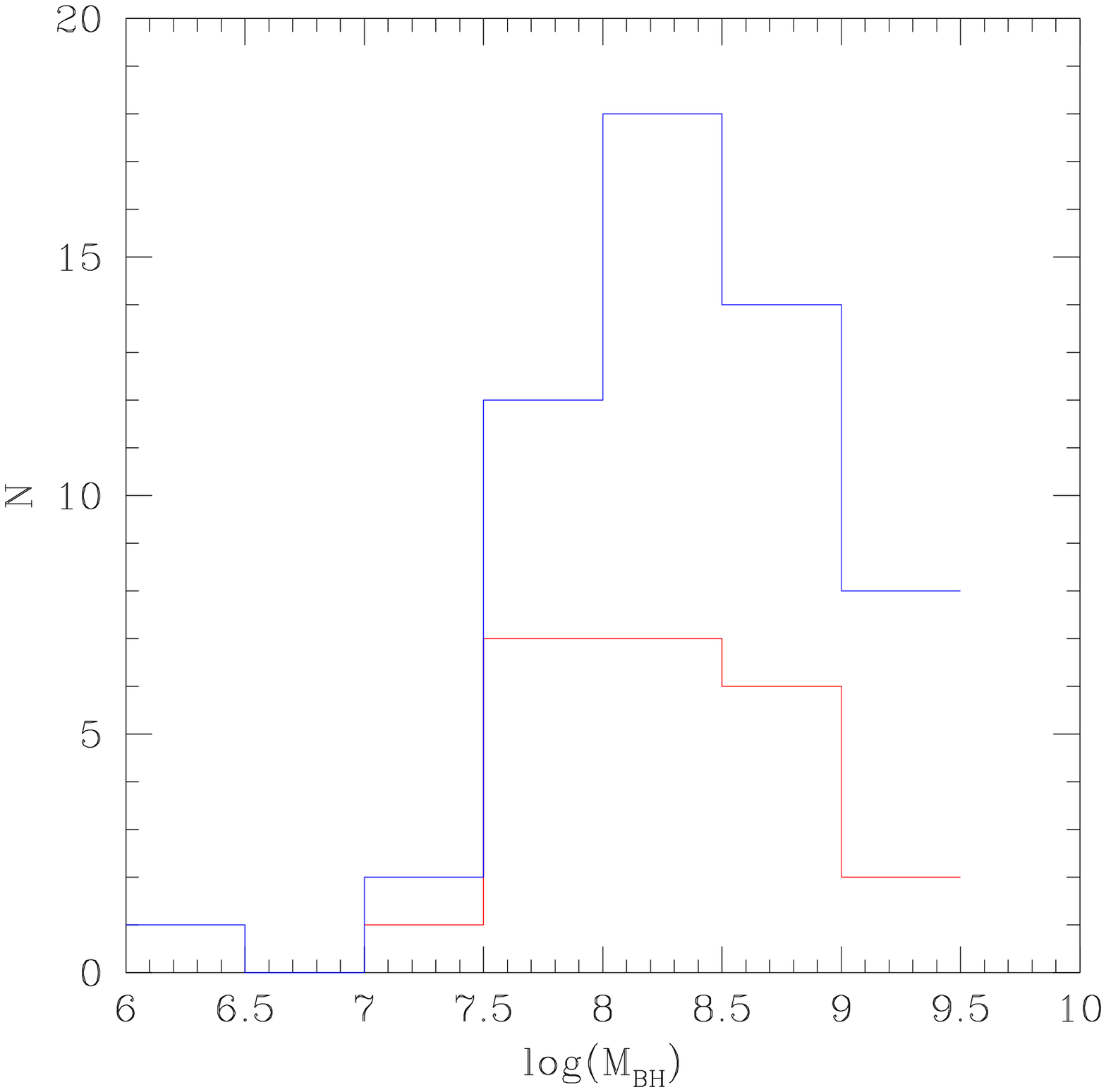}
\includegraphics[width=5cm, angle=0]{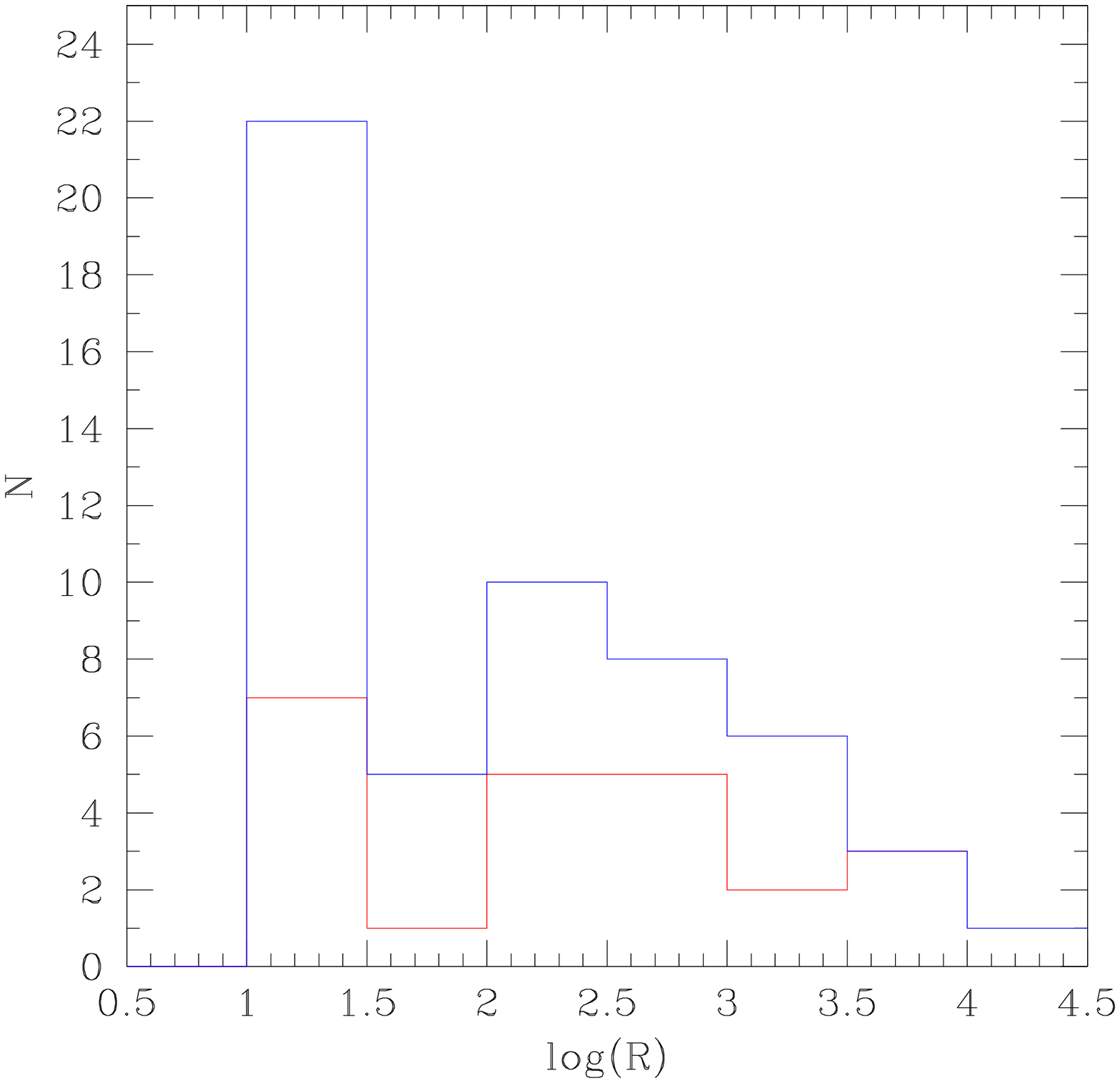}
\includegraphics[width=5cm, angle=0]{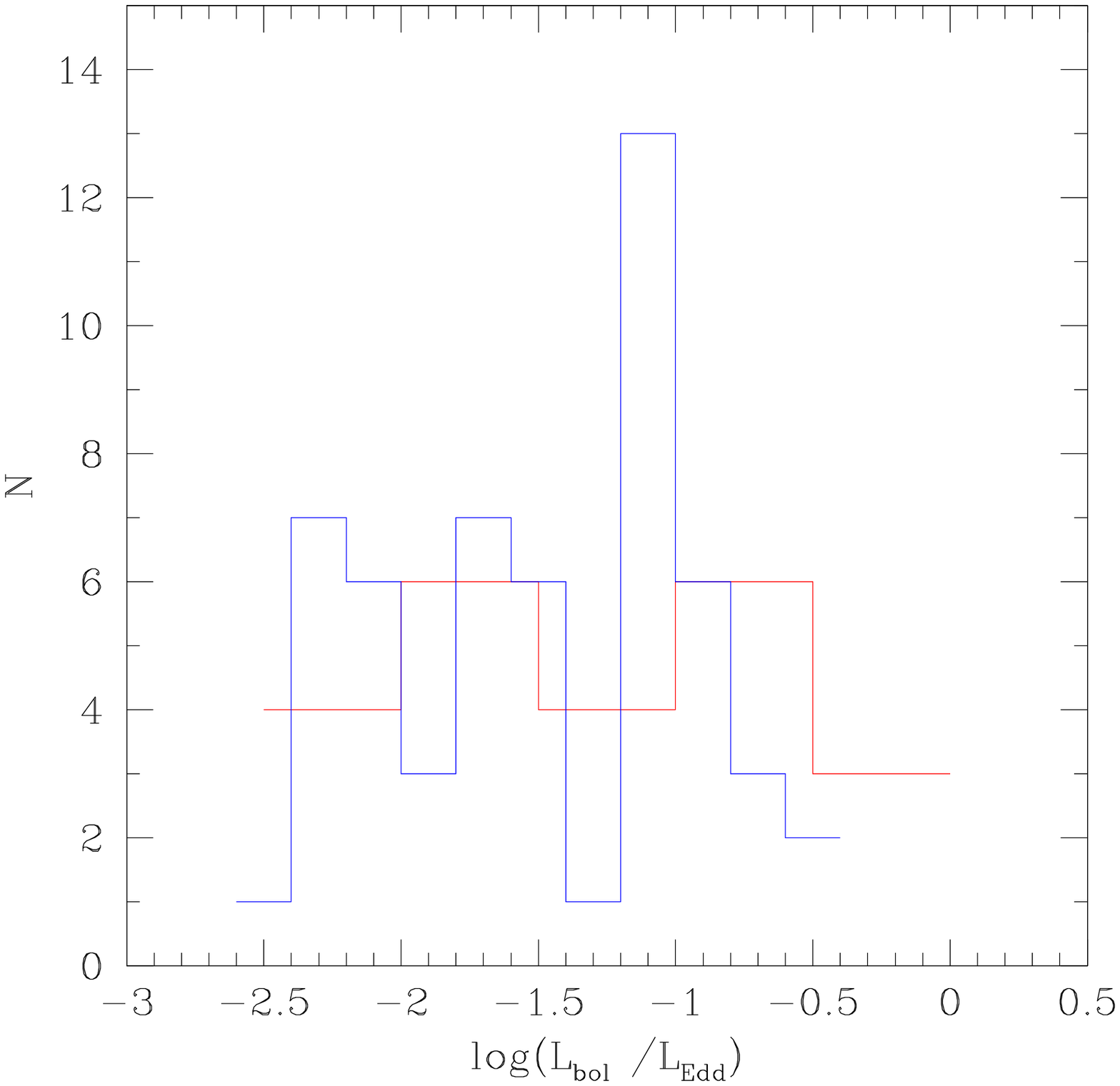}
\caption{Histograms of logarithms of sample parameters: top row for the entire sample, $M_{BH}$  (left); radio loudness (middle);
 Eddington ratio (right).  Bottom row for FSRQs (red) and SSRQs (blue): $M_{BH}$ (left); radio loudness (middle); and Eddington ratio (right)}.

\end{figure*}

\subsection{[O III] line properties}

As described earlier, [O III] profiles are modeled with two Gaussians, and the lower wavelength peaked Gaussian in both [O III]
lines represents wing component.
 We calculate the shift of [O III] core with respect to the rest frame wavelength, $v_c$ (in km $s^{-1}$).
  Some of the quasars have shown blueshifts  as well as redshifts of the [O III] lines. As defined in 
Komossa et al. (2008) for Narrow line Seyfert galaxies, a source is defined as a blue outlier (BO) when its velocity, $v_{[O III]} < 
-150$ km $s^{-1}$. In some of the quasars, [O III] lines are shifted towards higher wavelength and those are called red outliers. 
The measurement of velocity shifts of [O III] profiles should be done relative to the galaxy rest frame which is usually found
by measuring stellar absorption features. However, in our sample of AGNs, these are very weak or absent. The redshift is also
provided by the SDSS pipeline from which it is determined based on all strong emission lines. Hence, it could influence the
velocity shift of [O III] profiles. Therefore, we measure the redshift based on the narrow lines of $H_{\beta}$.
In our sample of radio loud quasars, 31 quasars are blue outliers whereas 7 of the quasars are red outliers.
Hence, around 17 \% of the quasars in our sample show outliers. In previous studies of Narrow Line
Seyfert 1 galaxies (NLS1s), Zamanov et al. (2002) and Komossa et al. (2008) found that they occur in between 4 -- 16 \%.
The outlier velocities attained by these quasars lie between $419$ to $-315$ km s$^{-1}$. 

Strong turbulence in the NLR possibly leads to such outliers. Also, radio luminosity might affect the gas kinematics in such 
a way that powerful relativistic jets can be linked with the outliers' velocites (Tadhunter et al. 2001; 
Nesvadba et al. 2008). Here, we investigate whether
the shift of [O III] is clearly connected to the accretion rate of radio loud quasars. We search for correlations of
$v_{c}$ with respect to $M_{\rm BH}$, L$_{rad}$, the Eddington Ratio and $L_{[O III]core}$; these are shown in Fig 5. The solid 
line represents the minimal value for outliers which is $-150$ km $s^{-1}$ as defined in Komossa et al. (2008) for NLS1s. It can 
be seen from the panels that no significant correlation is present among these quantities. 

The velocity of the wing component, $v_{w}$, is calculated with respect to the core component. All of
the quasars in our radio loud sample show blue wings with velocities up to 420 km $s^{-1}$.  A substantial subset (51) of the quasars
also show red wings with velocities up to $-316$ km $s^{-1}$.
We searched for a correlation between velocity of wing component with respect to the Eddington ratio, as these are thought to 
originate in outflows induced by high Eddington ratios (Komossa et al. 2008). Also, we test correlations of $v_{w}$ with respect to
$M_{\rm BH}$,  L$_{rad}$ and $FWHM_{[O III]core}$. Results are shown in Fig 6.  We do not find any 
significant correlation between these quantities. We then separated outliers from regular sources and searched if a correlation 
is present between these quantities. Only a weak correlation ($r$=0.236 with $p$=0.0092) was found between $v_{w}$ and
FWHM $[O III]_{c}$ for outliers in our sample, which indicates that a turbulent outflow is generated in the gas and
this turbulence results in a high FWHM [O III] core.

\begin{figure}
\centering
\includegraphics[width=4.0cm , angle=0]{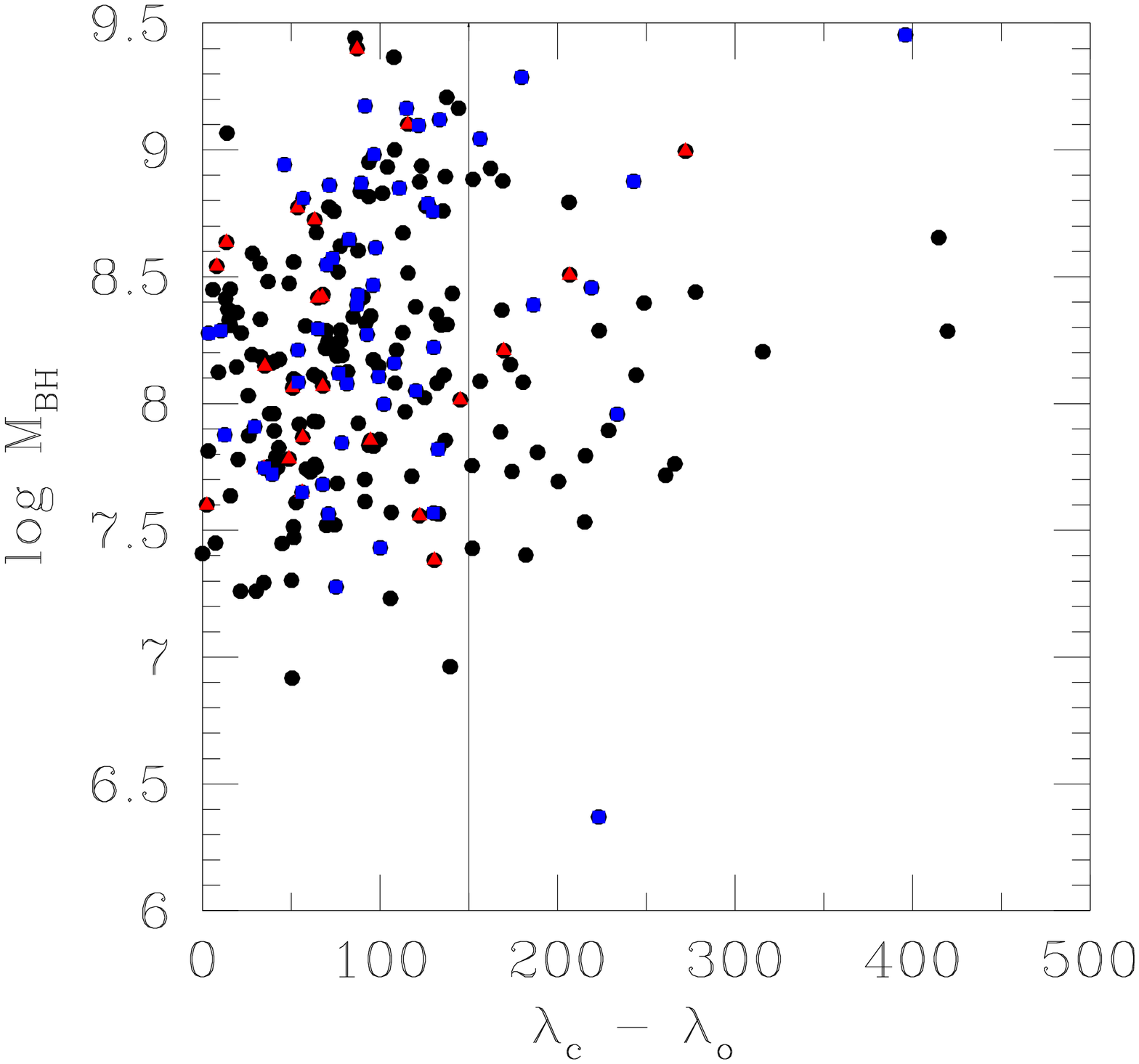}
\includegraphics[width=4.0cm , angle=0]{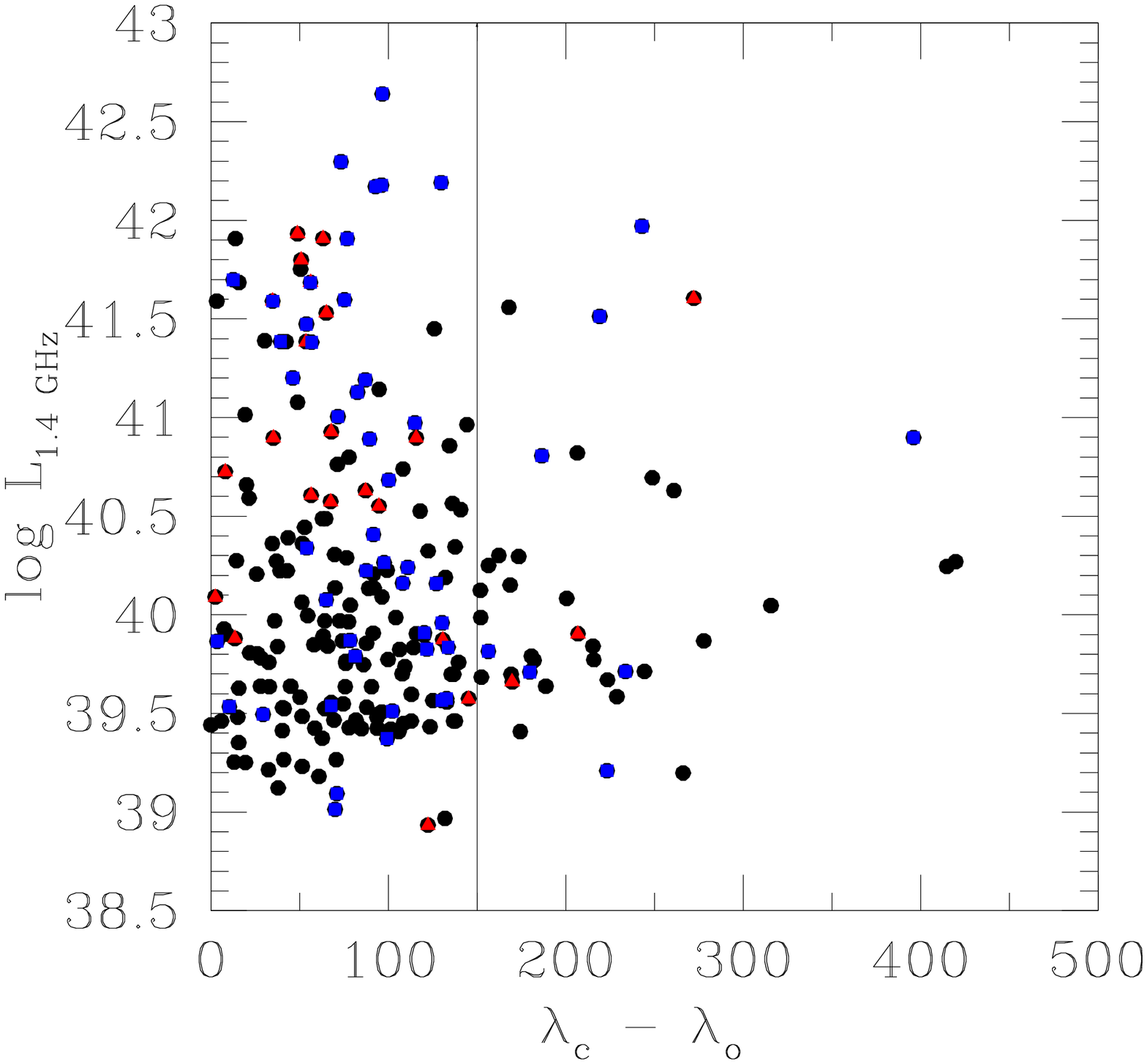}
\includegraphics[width=4.0cm , angle=0]{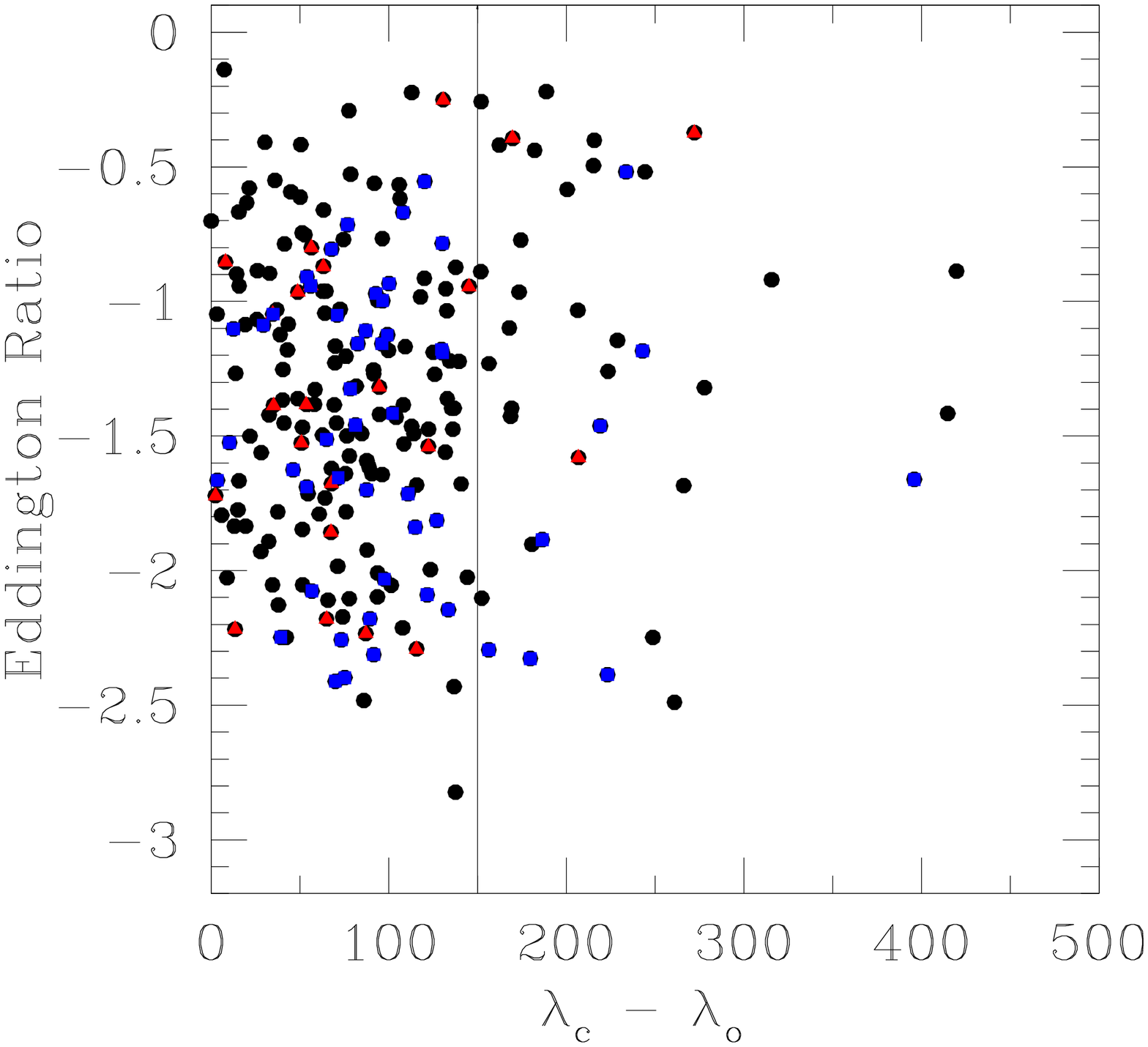}
\includegraphics[width=4.0cm , angle=0]{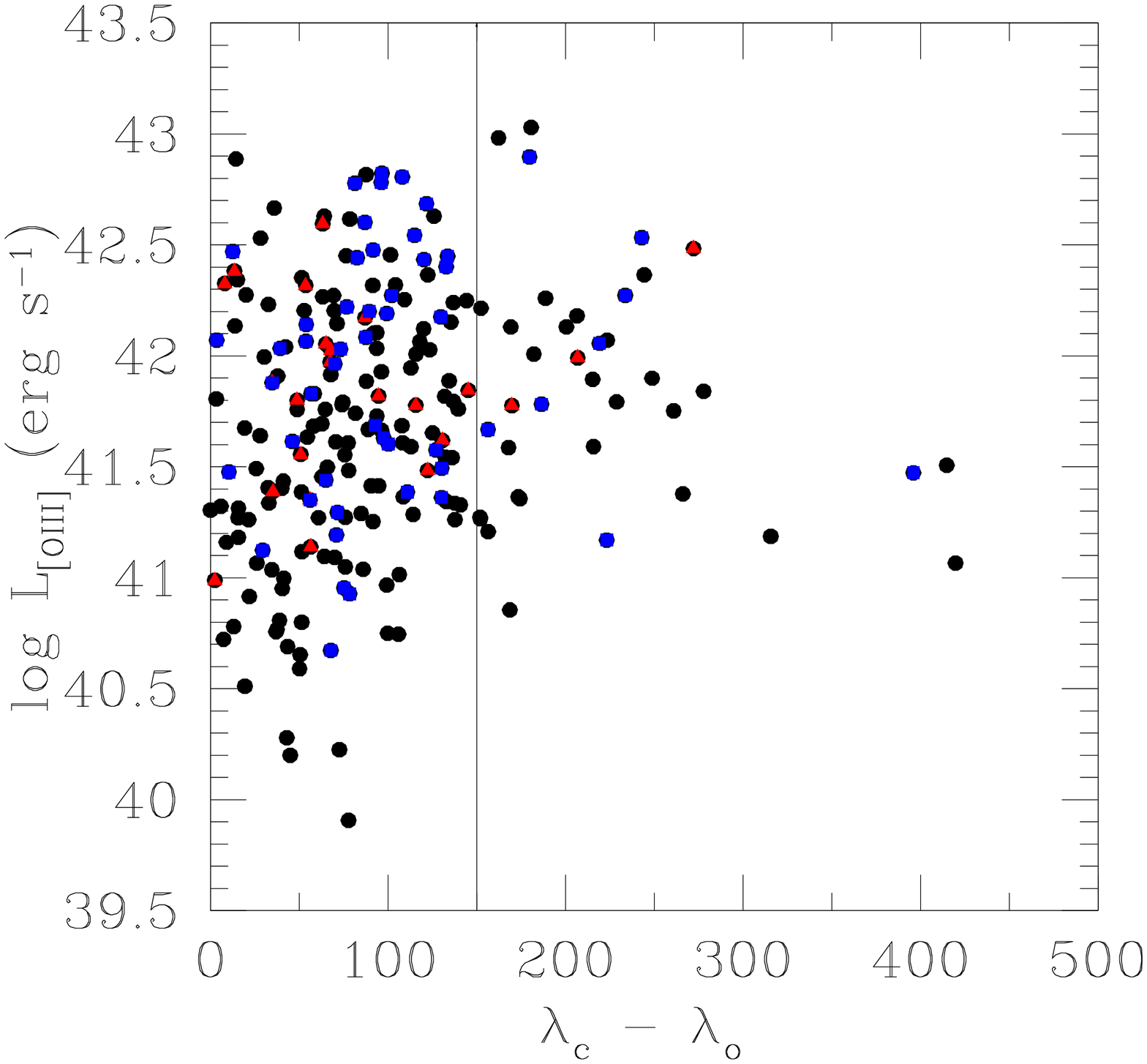}
\caption{Relation of absolute value of velocity of [O III] wing (in km/sec), in abscissa with respect to black hole mass (top left);
radio luminosity (top right); Eddington ratio (bottom left); and luminosity of $[OIII]^{c}$ (bottom right).
}
\end{figure}

\begin{figure}
\centering
\includegraphics[width=4.0cm , angle=0]{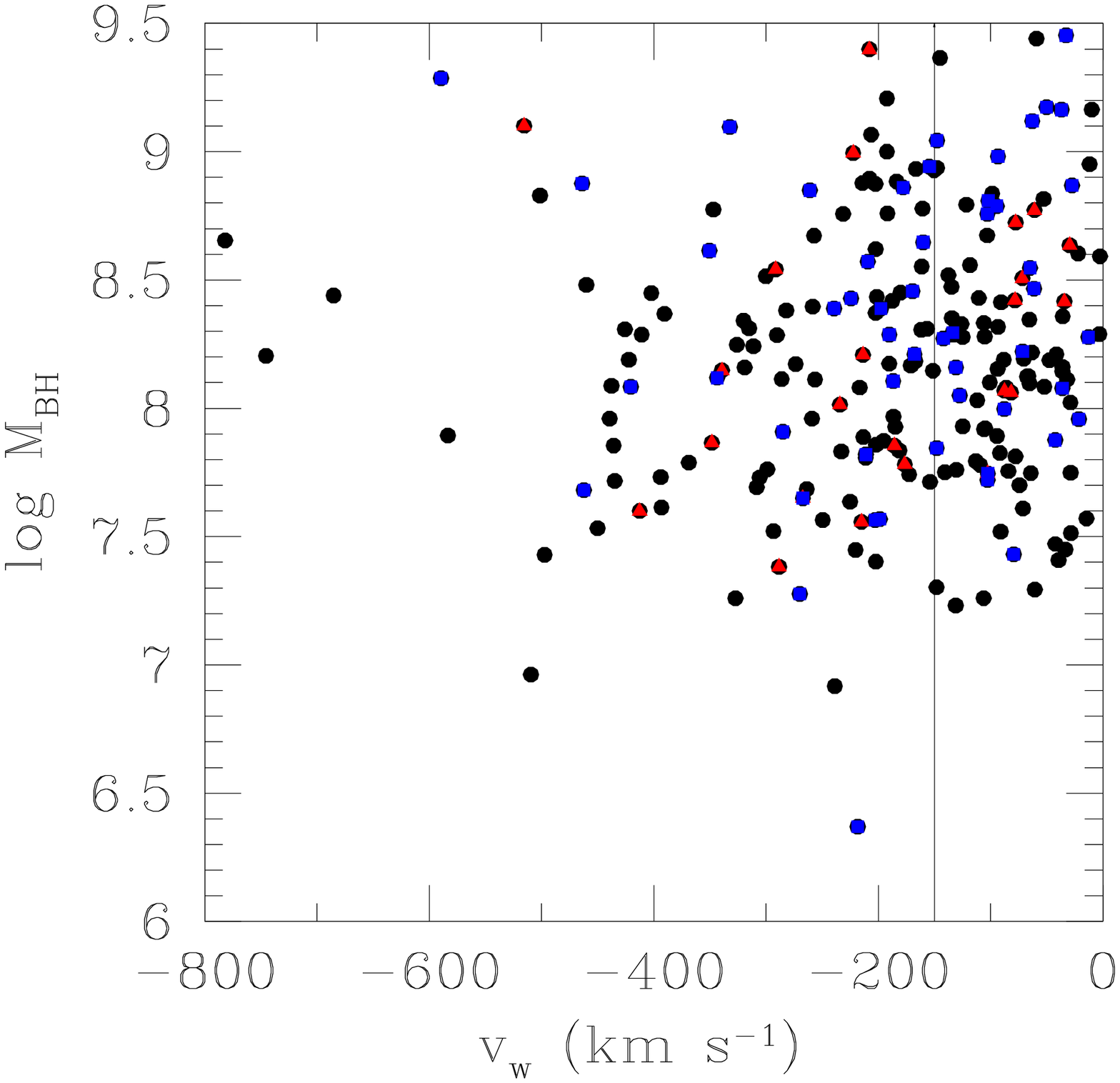}
\includegraphics[width=4.0cm , angle=0]{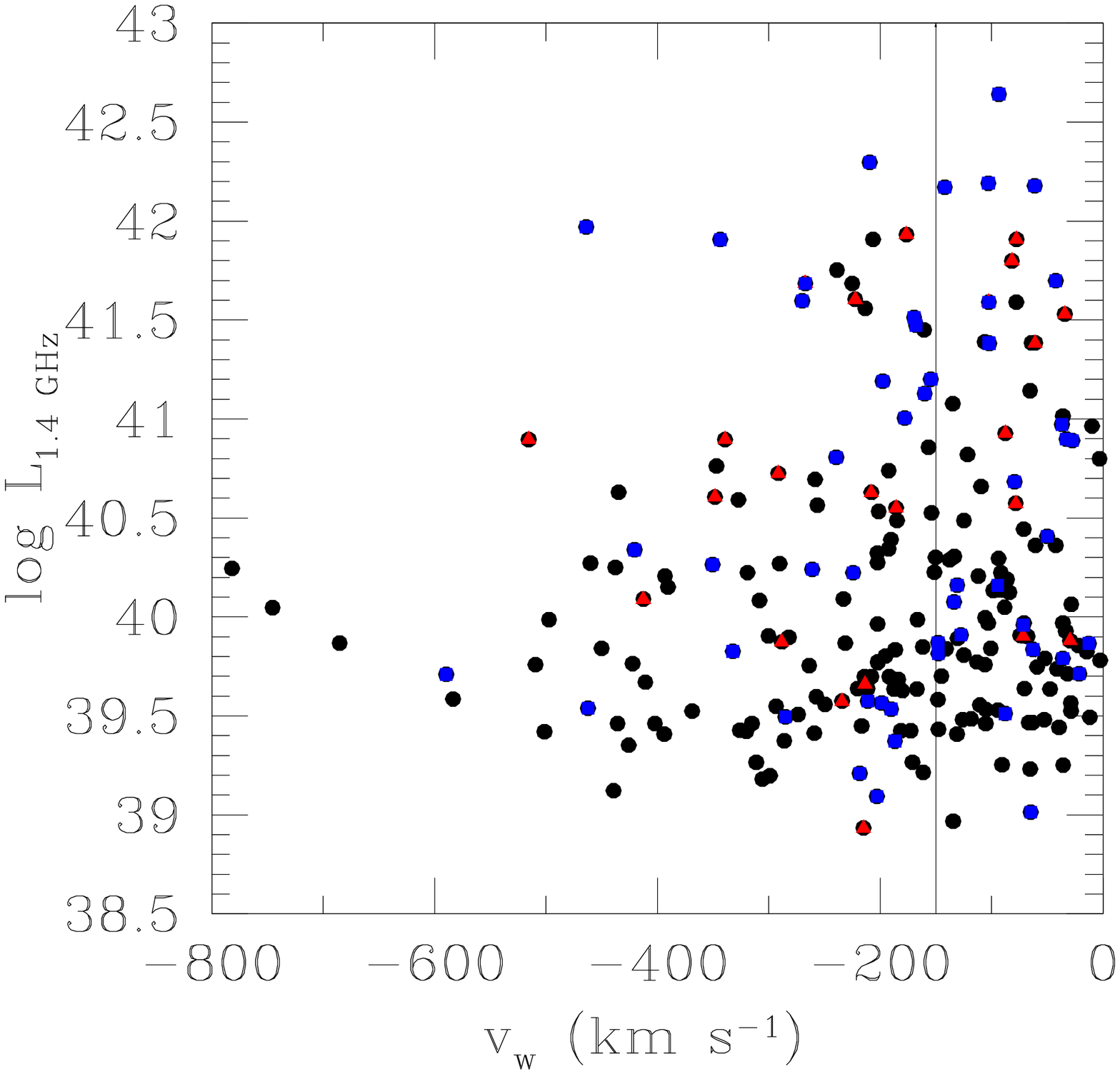}
\includegraphics[width=4.0cm , angle=0]{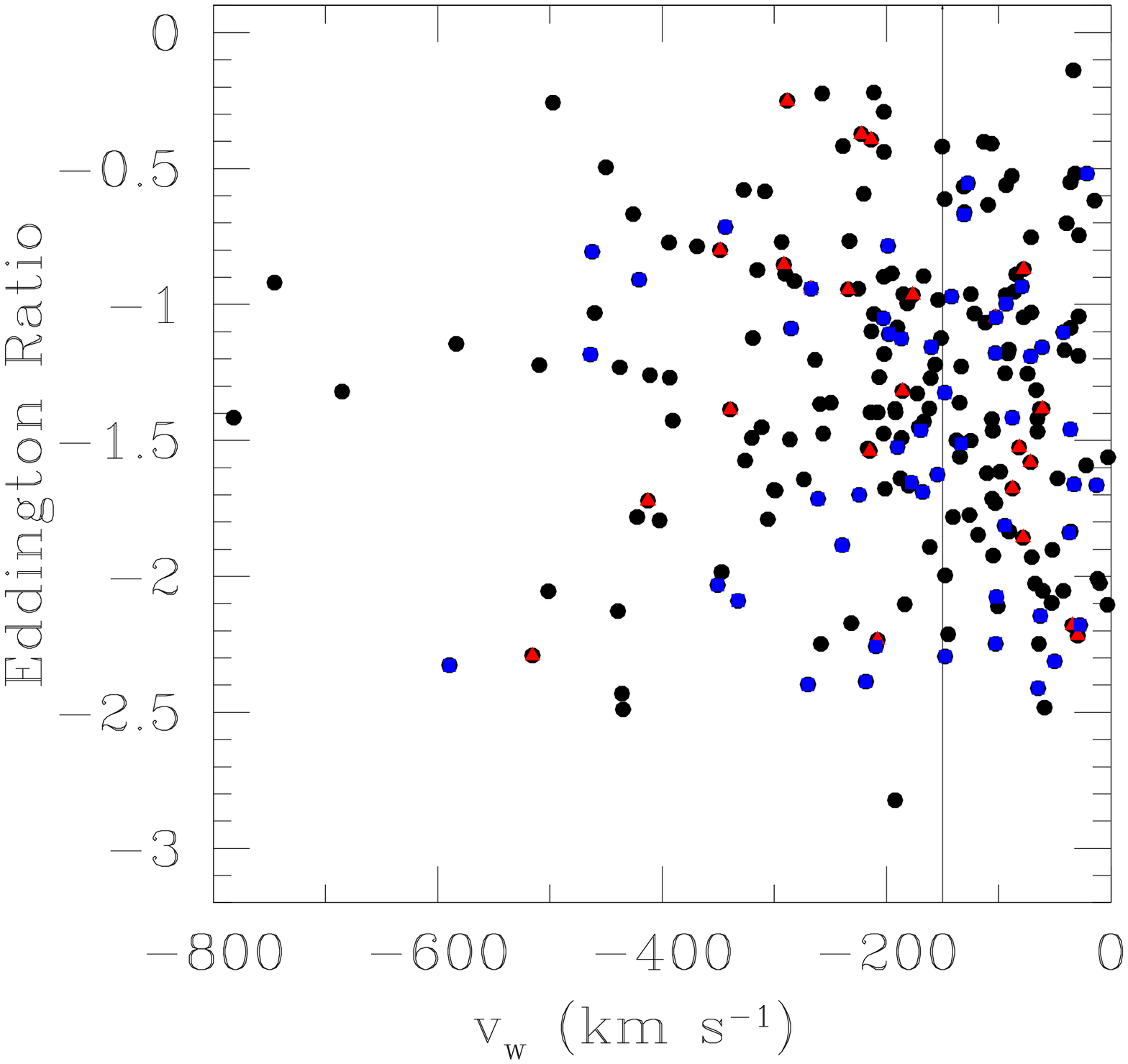}
\includegraphics[width=4.0cm , angle=0]{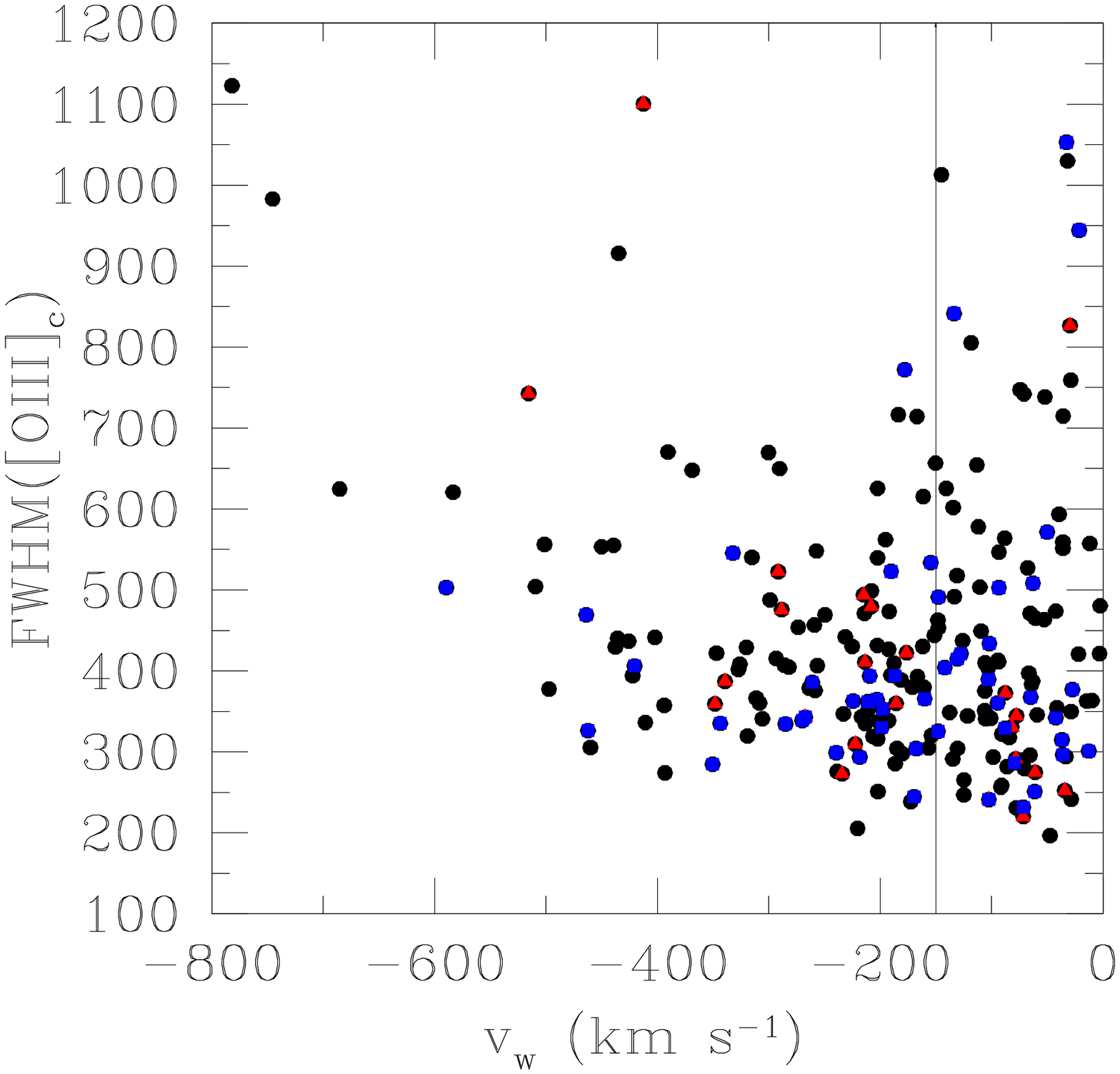}
\caption{Relation of absolute value of velocity shift of [OIII] (core), in abscissa with respect to black hole mass (top left);
radio luminosity (top right); Eddington ratio (bottom left) and FWHM $[OIII]^{c}$ (bottom right).
}
\end{figure}

\begin{figure}
\centering
\includegraphics[width=7.5cm , angle=0]{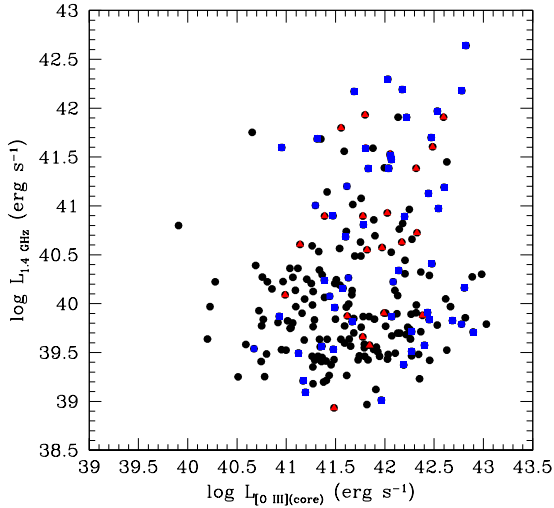}
\caption{Relation of radio luminosity, $log L_{\rm 1.4 GHz}$ (erg $s^{-1}$) to luminosity of [O III]$_{c}$ .
Color symbols are same as in figure 2.
}
\end{figure}

Also, we find a weak correlation between $\it L_{\rm 1.4 GHz}$ and the luminosity of the [O III] core, $L_{\rm [O III],core}$ 
with a Spearman correlation coefficient of $r_{\rm s}=0.253$ at confidence level $99.6\%$ (see Fig.\ 7); this indicates a possible
 relation between radio jets and NLR. 
 After removing the common dependence
on redshift using partial correlation analysis, we still find correlation value of $p_{\rm s}=0.225$ at confidence level 98.9\%. This could be explained by the shock 
excitation model where the NLR emission could be powered by radio-emitting jets (Bickell et al. 1997; Melendez et al. 2008). 
The flux limit of GB6 is much higher than that of FIRST, therefore, the source classification, especially those steep-spectrum
objects, is heavily biased towards objects with high flux density at 1.4 GHz. This probably explains the predominance of
unclassified objects at the low $L_{1.4GHz}$ side of Fig.\ 7.

It has been found in previous studies (Mullaney et al. 2013) that the radio luminosity has influence on the [O III] line profile. 
In order to investigate this effect, we performed a correlation analysis between $\it L_{\rm 1.4 GHz}$ 
and the FWHM of the core component, $FWHM_{\rm [O III],core}$. However, we do not find any significant correlation between these
quantities.
 
We also calculate the R5007 parameter which is defined as the ratio between [O III] $\lambda$5007 line and the whole $H_{\beta}$ flux.
Since, $H_{\beta}$ is formed in inner part of BLR and [O III] is formed in NLR, this ratio could be used to evaluate whether the jet 
interaction is different in BLR and NLR. 
We therefore searched for a possible correlation between R5007 and wing velocity, as fast [O III] wings lead to a reduction of
the covering factor in the NLR, which leads to a reduction of flux of [O III] lines. But for our sample of radio loud quasars
we do not find a significant correlation between these quantities.

\subsection{$\it M_{\rm BH}-\sigma$ relation}

We use the NLR gas velocity dispersions of [S II] and [O III] as surrogate for stellar velocity dispersion,
as follows, \begin{equation}\it \sigma = \sqrt{\it \sigma_{\rm obs}^{2}  - [\it \sigma_{\rm ins}/\rm (1+\it z)]^{2}}\end{equation} 
where $\it \sigma_{\rm obs}$= $\rm FWHM_{\rm [SII], or [OIII]}$/2.35 and $z$ is the redshift (Bian et al. 2008). For SDSS spectra, 
 mean value of the instrument resolution, $\it \sigma_{\rm ins}$, is about 56 km $\rm s^{-1}$ and 60 km $\rm s^{-1}$ for
 [S II] and [O III], respectively (Greene \& Ho 2005).

The $\it M_{\rm BH}-\sigma$ relation for our sample is shown in Fig. 8 using [S II] (upper panel)
and [O III] (lower panel) line width as surrogate for stellar velocity dispersion. Unclassified radio loud quasars, 
FSRQs and SSRQs are represented by black, red and blue symbols, respectively.
The Kormendy \& Ho (2013) relation for classical bulges/elliptical galaxies (KH13, Equation 1), the Tremaine et al. relation (2002)
 (T02) and the relation in McConnell \& Ma (2013) for late type galaxies are also shown in the 
figure using black, red and blue solid lines respectively. 

\begin{figure}
\centering
\includegraphics[width=9cm , angle=0]{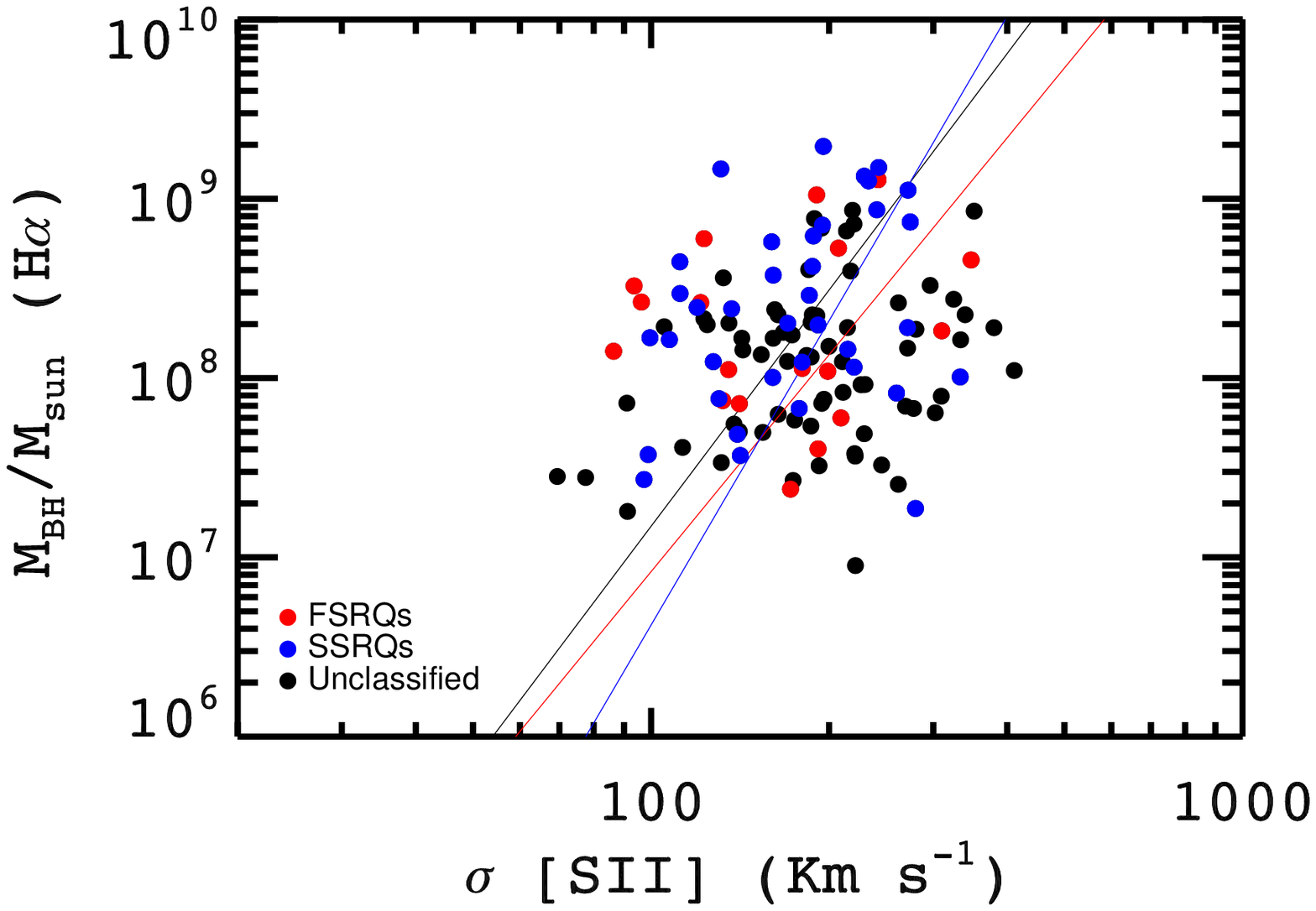}
\includegraphics[width=9cm , angle=0]{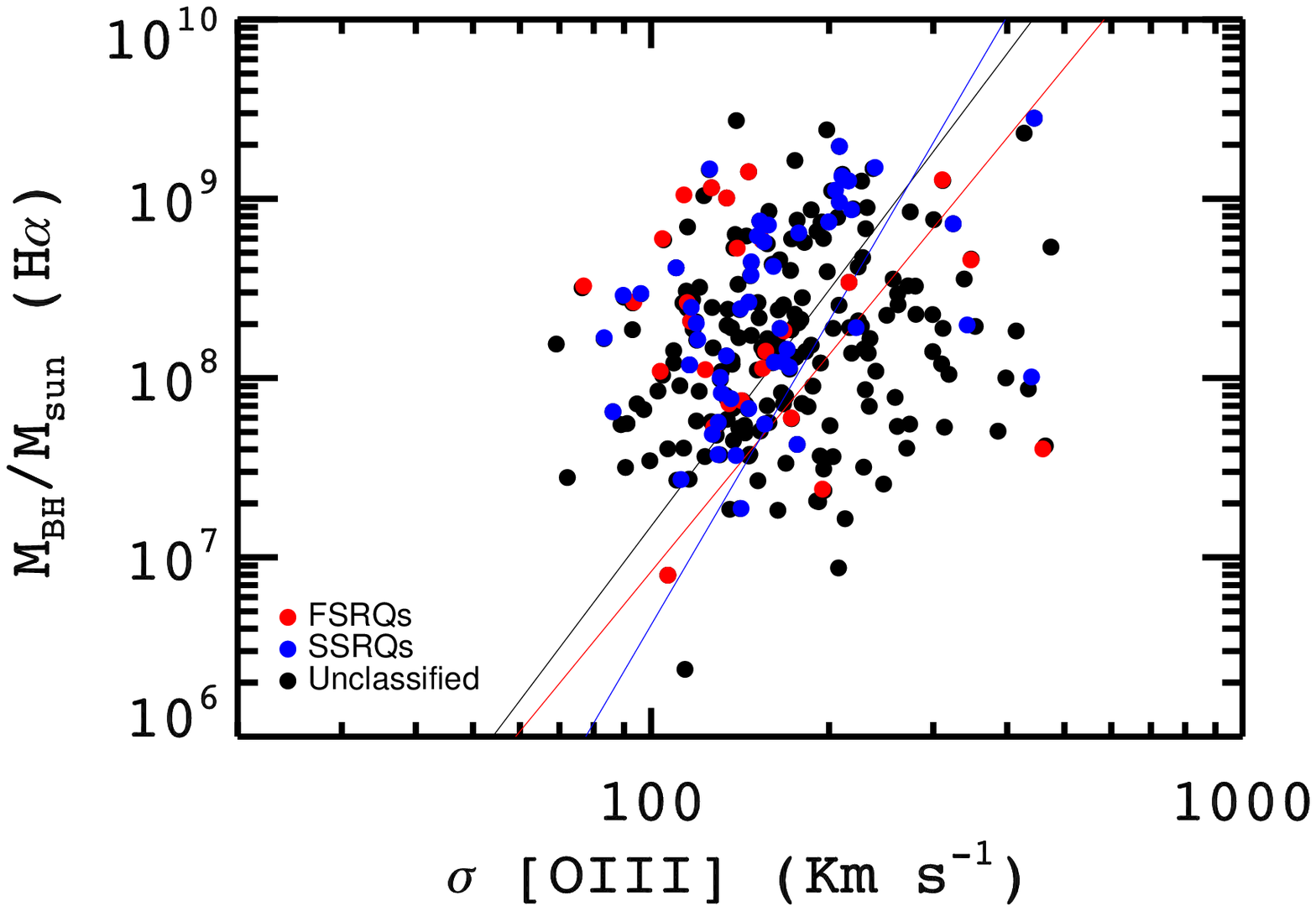}
\caption{$\it M_{\rm BH}$--$\it \sigma_{\rm [SII]}$ relation for radio-loud quasars (upper panel);
 FSRQs, SSRQs and Unclassified radio loud quasars are represented by red, blue and black symbols, respectively.
 $\it M_{\rm BH}$--$\it \sigma_{\rm [OIII]}$ relation of radio-loud quasars (lower panel). In both panels, 
black, red and blue solid lines represent the $\it M_{\rm BH}$--$\sigma$ relations for quiescent galaxies by Kormendy \& Ho (2013), 
Tremaine et al. (2002) and McConnell \& Ma (2013), respectively.
For [O III], the core of the line is used after decomposing asymmetric blue wings.} 
\end{figure}

\begin{figure}
\centering
\includegraphics[width=5.8cm , angle=0]{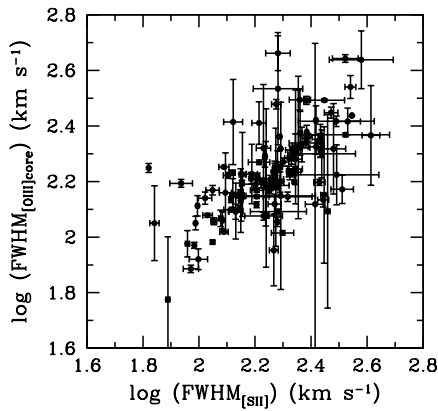}
\caption{ Correlations between the FWHM of [S II] and [O III], (Spearman Correlation coefficient, 
$r$=0.7527 with p=2.2e-16).}
\end{figure}

\begin{figure}
\centering
\includegraphics[width=9cm , angle=0]{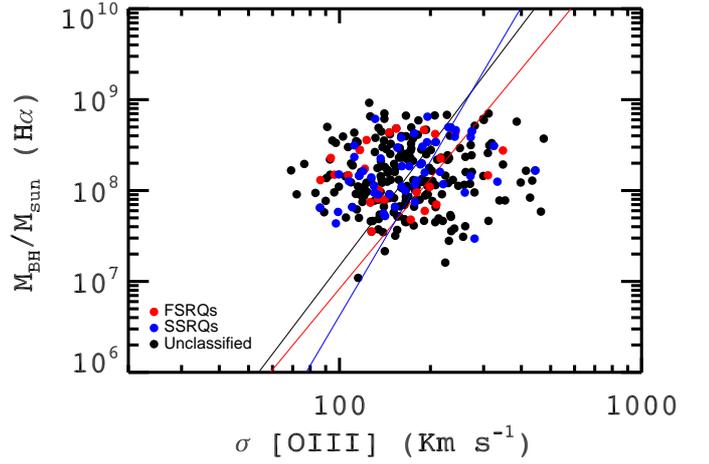}
\caption{$\it M_{\rm BH}$--$\it \sigma_{\rm [OIII]}$ relation of radio-loud quasars using the formalism of Wang et al. (2009)
and is given in equation (9). FSRQs, SSRQs and Unclassified radio loud quasars are represented by red, blue and black symbols, respectively.
Black, red and blue line represents the $\it M_{\rm BH}$--$\sigma$ relation of quiescent galaxies by Kormendy \& Ho (2013), Tremaine et al. (2002) and
McConnell \& Ma (2013), respectively.
}
\end{figure}

It is clear from the figure that the radio-loud quasars of our sample do not show any correlation between $\it M_{\rm BH}$ 
and $\sigma$,  although they do
cluster around the relation for local inactive galaxies, albeit with some outliers.

In order to find the overall offset of our sample of radio loud quasars from the local relation of inactive galaxies, we fit 
the log $\it M_{\rm BH}$ as a function of log $\sigma_{\ast}$:

\begin{equation}
\log (M_{\rm BH} / M_{\odot}) = \beta + \alpha \log (\sigma_{\ast} / 200\ \mathrm{km~s^{-1}}) .
\end{equation}
Here $y = \log (M_{\rm BH} / M_{\odot})$, $x = \log (\sigma_{\ast} / 200\ \mathrm{km~s^{-1}})$, $\alpha$ and $\beta$ are the slope and intercept
of the regression, respectively.
In order to perform the linear regression, we adopted two methods:  \texttt{FITEXY} (Tremaine et al. 2002) and  \texttt{LINMIX\_ERR} (Kelly 2007).

The \texttt{FITEXY} (Press et al. 1992), modified by Tremaine et al. (2002), is implemented in our work in IDL
using the \texttt{mpfit} (Markwardt 2009) Levenberg-Marquardt least-squares minimization routine.
Note that our implementation is basically similar to that provided in Williams et al. (2010)
\footnote{http://purl.org/mike/mpfitexy}.
It performs the linear regression by minimizing
\begin{equation}\label{method:chi2}
{\chi ^2} = \sum\limits_{i = 1}^N {\frac{{{{\left( {{y_i} - \alpha  - \beta {x_i}} \right)}^2}}}{{\sigma _{y,i}^2 + {\beta ^2}\sigma _{x,i}^2 + \epsilon _{{\mathop{\rm int}} }^2}}},     
\end{equation}
where $\alpha$ and $\beta$ are the regression coefficients,
$\sigma_{x}$ and $\sigma_{y}$ are standard deviation in measurement errors,
and $\epsilon^2_{{\mathop{\rm int}}}$ is the intrinsic variance. The value of $\epsilon_{{\mathop{\rm int}}}$ is iteratively 
adjusted as an effective additional $y$ error
by repeating the fit until one obtains $\chi^2/(N-2)=1$
(i.e., following the suggested iterative procedure given in Bedregal et al. 2006 and Bamford et al. 2006).
If after the initial iteration, the reduced $\chi^2$ is less than one, then no further iterations
occur and one sets $\epsilon_{{\mathop{\rm int}}}=0$.
$\epsilon$ properly accounts for the intrinsic scatter thus the best-fit slope is not biased due to a few points
with small measurement errors (e.g., see the discussion in Tremaine et al.2002). We only consider the measurement errors of 
log $M_{\rm BH}$ and $\sigma$ in the fit and thus the intrinsic scatter $\epsilon$
includes the contribution from the systematic errors in log $M_{\rm BH}$ and $\sigma$. We also tried adding the 
systematic errors on both  log $M_{\rm BH}$ and $\sigma$ but did not find significantly different results of slopes 
and intercept (also found in Shen et al. 2015).

We do not find any correlation between $\it M_{\rm BH}$ and $\sigma$ likely due to systematic uncertainties on both 
parameters, but there is clustering of points around the local relation.
In order to find the overall offset of our sample of radio loud quasars with respect to the local relation, we follow
the approach of Sheinis \& Lopez-Sanchez (2017) and fix $\alpha$ (to the slope of KH13) in our regressions.
The results of the regression are summarized in Table 1.
We estimate our results with the alternative regression method which uses the Bayesian linear regression routine, 
\texttt{linmix\_err}, developed by Kelly (2007). It is available in the NASA IDL astronomy user's 
library\footnote{http://idlastro.gsfc.nasa.gov/}.
We do not find significantly different results for slope and intercept using these two methods and therefore quoting results
of regression from the previous one.

In this regression analysis, we fix $\alpha = 4.38$  (the slope of the KH13 relation) and 
leave the intercept as a free parameter. The results of this regression for $\sigma$ of [S II], [O III] and for various redshift bins 
are provided in Table 1. 
It can be seen that while using [SII] lines as $\sigma_{*}$, the radio loud sample shows intrinsic scatter of 0.74 
around the local relation with intercept of 8.41. But, the sample consisting of $\sigma_{[SII]}$ involves selection bias 
because we removed most broader $H_\alpha$ profiles where [S II] lines were blended. Hence, for such quasars, [S II] line
 widths could not be calculated and are removed from the sample.
For the sample consisting of [OIII] line width, intrinsic scatter decreases to $\sim$0.60 with an overall deviation of 
$\sim$0.12 $\pm$0.05 dex from the local relation. However, if we take into account the systematic uncertainties
on $\it M_{\rm BH}$ and $\sigma$, these deviations would not be significant. 
Our results are in accordance with the previous studies where it has been found that  $\it M_{\rm BH}-\sigma$ relation 
for active galaxies appear to be flatter than for quiescent galaxies
(Woo et al. 2013; Shen et al. 2015). It is argued in Shen et al. (2015) that the flattening of the $\it M_{\rm BH}-\sigma$ relation
of active galaxies at high redshift is due to the various selection biases which lead to the sample containing more
luminous and massive systems. This shifts the most massive black holes  above the quiescent galaxy relation
(Woo et al. 2015; Salviander \& Shields 2013; Brotherton et al. 2015) and hence the $\it M_{\rm BH}-\sigma$ relation for quasars
in individual samples may change with redshift. 

In order to check the offset of our sample of $\it M_{\rm BH}-\sigma$ relation from quiescent galaxies at different 
redshifts, we divided whole sample into three redshift bins i.e. 0.1--0.2, 0.2--0.25 and 0.25--0.3 in such a way that 
nearly equal numbers of quasars fall in each bin. It can be seen from the parameters in Table 1 that there is 
significant offset  
 of $\sim$0.33$\pm$ 0.06 dex for the quasars lie in highest redshift bin (0.25--0.3). Intrinsic scatter 
decreases as we move from lowest to highest bin.

\begin{table}
\caption{Linear Regression results for the radio loud sample with $\alpha$ fixed at 4.38 as discussed in text.}
\begin{tabular}{ccc} \hline
  Sample              &$\beta$          &$\epsilon$\\ \hline
 full ([SII])          &8.412$\pm$0.069  &0.739   \\
 full ([OIII])(123)    &8.555$\pm$0.068  &0.706   \\
 full ([OIII])(223)    &8.610$\pm$0.050  &0.599   \\
0.1$<$z $<$0.2(62)     &8.546$\pm$0.108  &0.761    \\
0.2$<$z $<$0.25(80)    &8.516$\pm$0.082  &0.591    \\
0.25$<$z $<$0.3(80)    &8.814$\pm$0.062  &0.375    \\
 FSRQs                 &8.819$\pm$0.154  &0.569   \\
 SSRQs                 &8.741$\pm$0.082  &0.404    \\ \hline
\end{tabular}
\end{table}

\subsubsection{Comparison of $\it M_{\rm BH}$ with alternate method}

In Section 3.1, we used the empirical relation of Reines, Greene \& Geha (2013) to calculate the virial black hole mass.
As mentioned in section 3.1, it is based on the methodology outlined in Greene \& Ho (2005), but derived with  
modified radius-luminosity relationship of Bentz et al. (2013). As compared to $\it M_{\rm BH}$ estimation using equation (4) of
 Greene \& Ho (2005), this modification of radius-luminosity relation causes an increase in the estimation of
$\it M_{\rm BH}$ by a factor of $\sim$ 1.6. Estimated black hole mass range lies between $\it M_{\rm BH}$=
 10$^{7.1}$--10$^{9.2}$ $\it M_{\rm \odot}$ (in section 4.1). 

Wang et al. (2009) presented a new formalism of the empirical relation using $\rm H\beta$ lines and the updated 
BH mass measurements from reverberation mapping.
This new formalism has shown improved internal scatter between the single-epoch estimators and the mass estimators 
based on reverberation mapping,
but systematically deviates from some of the commonly used $\it M_{\rm BH}$ estimators in the literature.
It involves $\it M_{\rm BH}$ $\propto$ $\rm FWHM^{1.09}$ instead of  $\it M_{\rm BH}$ $\propto$ $\rm FWHM^{2}$ commonly used 
in the literature, which
gives progressively higher and lower  $\it M_{\rm BH}$ values towards the low and high mass ends, respectively. Collin et al. (2006) also gives results consistent with those that of the Wang et al. (2009) formalism over a relatively large mass range. Hence, the discrepancy
between previously used mass estimators and Wang et al. (2009) arises because of the use of more recently recalibrated and updated reverberation
mapping $\it M_{\rm BH}$ measurements from literature (i.e. Peterson et al. 2004; Bentz et al. 2007; Grier et al. 2008) and use of best fitting
value of $\gamma$ instead of fixing $\gamma$ =2 in  $\it M_{\rm BH}$ $\propto$ $\rm FWHM^{\gamma}$.
In order to check whether this systematic bias described by Wang et al. (2009) can affect the intrinsic scatter in 
$\it M_{\rm BH}$ estimation of our radio-loud sample, we re-estimated the $\it M_{\rm BH}$ of our sample using the formalism of 
Wang et al. (2009), which is as follows:

\begin{equation}
\begin{array}{l}

\rm log(\frac {\it M_{\rm BH}}{{\it M}_{\odot}}) = 7.39+0.5~ log\Big(\frac{{\it L}(5100)}{10^{44}\,erg\,s^{-1}}\Big)
+1.09~ log\Big(\frac{{\rm FWHM}(H\beta)}{10^3\,km\,s^{-1}}\Big).

\end{array}
\end{equation}

 The use of line luminosity and FWHM  of $\rm H\alpha$ are better in estimating $\it M_{\rm BH}$ of radio loud quasars
(due to the reasons mentioned in section 3.1). Therefore, to transform equation (9) into the line luminosity and 
FWHM of $\rm H\alpha$, we used the empirical relations provided by Greene \& Ho (2005) between the FWHM of $\rm H\alpha$ 
and $\rm H\beta$, and between the broad $\rm H\alpha$ luminosity and continuum luminosity at 5100 \AA, $L_{\rm 5100}$ 
and put them in equation 9.

\begin{equation}
\begin{array}{l}

\rm FWHM (H\beta) = 1.07 \times 10^{3} {(\frac{FWHM (H\alpha)}{ 10^{3} km~ s^{-1}}})^{1.03} ~km ~s^{-1}\\

\end{array}
\end{equation}

\begin{equation}
\begin{array}{l}

\it L_{\rm H\alpha}=\rm 5.25 \times 10^{42} {(\frac{\it L_{\rm 5100}}{10^{44} \rm erg ~s^{-1}}})^{1.157} \rm erg ~s^{-1}
\end{array}
\end{equation}

By putting the above relations in Equation (9), we get the formula as below:

\begin{equation}
\begin{array}{l}

\rm log(\frac {\it M_{\rm BH}}{{\it M}_{\odot}}) = 7.11 + 0.43~log\Big(\frac{{\it L}(H\alpha)}{10^{42}\,erg\,s^{-1}}\Big)
+1.12~log\Big(\frac{{\rm FWHM}(H\alpha)}{10^3\,km\,s^{-1}}\Big)

\end{array}
\end{equation}

Now we reestimate $\it M_{\rm BH}$ using equation (12) and we replot the $\it M_{\rm BH}-\sigma$ relation using these
newly estimated  $\it M_{\rm BH}$ values in Fig.\ 10.   The $\it M_{\rm BH}$ range is slightly 
reduced to 10$^{7.1}$--10$^{8.8}$ $\it M_{\rm \odot}$ . 
It can be seen that the scatter is indeed reduced, as pointed out by Wang et al. (2009). However, we find that the 
newly estimated $\it M_{\rm BH}$ values do not change our main results.

\subsubsection{Biases and uncertainties}

The intrinsic scatter along vertical direction in the $\it M_{\rm BH}-\sigma$ relation of radio loud quasars
 can be partly accounted by the uncertainties in estimation of the
black hole masses, which can be under- or over-estimated due to the BLR geometry and Doppler boosting, respectively.
The Doppler boosting effect has been avoided by replacing continuum luminosity with the line luminosity in BLR
radius-luminosity empirical relation. Similar black hole mass distributions between FSRQs and SSRQs shows that the
BLR geometry effect may not be severe in our sample. Besides these two effects, the single-epoch virial BH mass estimators
are still subject to a number of uncertainties that are propagated from the measurement errors
in FWHM and line luminosity and different methods adopted to estimate line widths and luminosities, leading to various discrepancies
(Shen et al. 2008); however, the dominant uncertainty is the systematic uncertainty which is $\sim$0.5 dex (Shen 2013).

Uncertainty is also present in the estimation of $\sigma$. As pointed out earlier that the FWHM of [S II] and [O III] 
emission lines are used as a surrogate for stellar velocity dispersion $\sigma$. These lines are used as $ \sigma$  
in previous studies to explore $M_{\rm BH}-\sigma$ relation  (i.e., Gu et al. 2009; Salviander \& Shields 2007; 2013; Brotherton
et al. 2015 and references therein), finding large scatter. The uncertainty of this substitution is large as  shown by the direct
comparison between [O III] line width and the stellar $ \sigma$ (Xiao et al. 2011; Woo et al. 2016; Sheinis et al. 2017). For type 2 
AGNS, Woo et al. (2016) found an uncertainty of 0.19 dex in the direct comparison of [O III] line width and $\sigma$. 
The actual uncertainty for the above substitution is not known for radio-loud quasars, hence, we are assuming a lower 
limit of 0.19 dex in the substitution of $\it \sigma_{\rm [OIII]}$.
Along with these uncertainties, the estimation of $\it \sigma_{\rm [OIII]}$ can also be influenced by the outflows which can be 
noticed through a moderate correlation between the  $\it \sigma_{\rm [OIII]}$ and the velocity width of the outflowing gas 
(shown in Fig. 11). 

Now we discuss the other potential biases (apart from uncertainties in the measurements of $\it M_{\rm BH}$ and $\it \sigma$)
 that could lead to the observed intrinsic scatter in our sample of radio loud quasars. This includes the intrinsic scatter 
in the $\it M_{\rm BH}-\sigma$ relation of
inactive galaxies of 0.31 dex for early type galaxies and 0.44 dex for all galaxy types, based on the locally observed sample of galaxies
(Gultekin et al. 2009). However, the magnitude of intrinsic scatter in the $\it M_{\rm BH}-\sigma$ relation of AGN samples are not known.

Salviander et al. (2007) simulated the effect of Malmquist-like bias which arises from correlations between quasar luminosity,
$\it M_{\rm BH}$ and redshift. Also, Lauer et al. (2007) suggested that because there is intrinsic scatter in the
 $\it M_{\rm BH}-\sigma$ relation, the samples selected based on a threshold in quasar luminosity will preferentially select
over-massive BHs with respect to the stellar velocity dispersion. Shen \& Kelly (2010) and Shen (2013) found that because
single epoch virial mass estimates depend on luminosity, quasar samples with high threshold luminosity are biased towards high
virial masses. Shen \& Kelly estimate a bias of 0.2$-$0.3 dex in $\it M_{\rm BH}$ for $L_{\rm bol}$ $>$ $10^{46}$ erg $s^{-1}$.
Recently, Shen et al. (2015) performed simulations to show that these biases lead to a flattening in the slope of
$\it M_{\rm BH}-\sigma$ relation as the under-massive BHs are more easily lost due to their high luminosity threshold.
 The effects of the above mentioned statistical biases are more severe at high redshifts.

\begin{figure}
\centering
\includegraphics[width=7.2cm , angle=0]{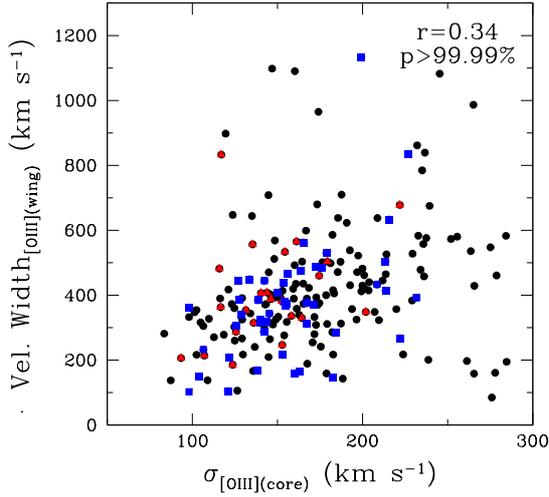}
\caption{Relation of $\it \sigma_{\rm [OIII]}$ w.r.t the velocity of the outflowing gas, $\it V_{\rm [O III] wing}$.
Representation of color symbols are same as in figure 4.
}
\end{figure}

\section{Discussion \& Conclusions}
 
In this work, we studied a sample of 223 radio loud quasars out to $z < 0.3$. We investigate their black
hole mass, radio luminosity and Eddington ratio distributions.
We calculated the radio spectral indices and were able to classify 26 of the quasars as FSRQs and 56 as SSRQs.

We investigate the [O III] properties of our radio loud sample and find that around 17\% of these quasars 
show outliers. The typical velocity attained by these outliers lies between 419 to $-$315 km s$^{-1}$. These outliers are 
thought to originate either from strong turbulence in the NLR or the influence of powerful 
relativistic jets. But, we do not find any significant correlation of velocity shift of [O III] with respect to
$L_{1.4GHz}$, Eddington ratio, etc. Also, all the [O III] profiles have shown blue wings with velocities up to
420 km s$^{-1}$. Blue wings are also believed to originate in outflows that are induced by high Eddington
ratio but we do not find any significant correlation between these quantities. 

Correlation between $L_{1.4GHz}$ and $L_{[O III] core}$ indicates a possible relation between radio jets
and NLRs. We find only a weak correlation between these quantities. 

It is found that quasars having broader $\it \sigma_{\rm [OIII core]}$ are associated with high outflow velocity fields. 
Hence, the effect of outflows could influence the true estimation of $\it \sigma_{\rm [OIII]}$
by broadening the [O III] line profiles (Dopita \& Sutherland 1995; Nelson \& Whittle 1996; Komossa \& Xu 2007).

We revisit the $\it M_{\rm BH}-\sigma$ relation for our sample of radio-loud SDSS quasars using [OIII] and [SII] $\lambda$6716, 6731
$\sigma$ as surrogate for stellar velocity dispersion. We find that the radio-loud
quasars do not show a relationship between $\it M_{\rm BH}$ and $\sigma$  which is expected due to the systematic uncertainties
on them (mentioned above) but instead cluster around the $\it M_{\rm BH}-\sigma$
relation for inactive galaxies.

Some recent works have shown that the $\it M_{\rm BH}-\sigma$ relation for active galaxies appear to be
shallower/flatter than for inactive/quiescent galaxies (Woo et al. 2013; Shen et al. 2015). 
Both the studies involve measurement of stellar absorption features to calculate the stellar velocity dispersion. Woo et al. (2013) found
different slopes of $\it M_{\rm BH}-\sigma$ relation for quiescent and active galaxies which could be due to the real physical difference of 
BH-galaxy coevolution. However, Shen et al. (2015) also found flattening of the $\it M_{\rm BH}-\sigma$ relation of active galaxies at high redshift
but they argued it could be due to the various selection biases and BH mass uncertainties in luminosity-threshold quasar samples.
As the redshift increases, most of the selection criteria lead to the sample containing more luminous and massive systems
which shifts the most massive black holes above the quiescent galaxy relations (Woo et al. 2010; 2013, Salviander \& Shields 2013, Shen 
et al. 2015; Brotherton et al. 2015) and hence the $\it M_{\rm BH}-\sigma$ relation of quasars in individual samples may change with 
redshift.
Our sample is restricted to z $\leq$ 0.3 since [S II] was used. We find an overall offset of 0.12$\pm$0.05 of our sample of 
radio loud quasars from the local relation of quiescent galaxies. 
When the quasars in highest redshift bin (0.22--0.3) are considered, an 
overall offset of 0.33$\pm$0.06 is found.  These results are used with caution as systematic uncertainties are not taken into
account while calculating them.
 A detailed study of the $\it M_{\rm BH}-\sigma$ relation of radio-loud quasars 
at higher redshift is required, which will be done in our future work.
\\
We are grateful to the anonymous referee for his/her insightful comments.
We are indebted to Prof. Gregory Shields and Prof. Paul J. Wiita for carefully reading the manuscript and providing his
 very valuable suggestions on it. 
We are grateful to Prof. Arun Mangalam and Prof. J. H. Woo for their useful discussions and comments on the work. HG acknowledges 
Dr. Xiaobo Dong, Dr. Ting Xiao for helpful discussions on analysis and data reduction. HG thanks Liao Mai for help in this work. 
HG is sponsored by the Chinese Academy of Sciences Visiting Fellowship for Researchers from Developing Countries, CAS President's
 International Fellowship Initiative (grant No. 2014FFJB0005), supported by the NSFC Research Fund for International 
Young Scientists (grant 11450110398) and supported by the China Postdoctoral Science Foundation Grant (grant 2016T90393).
HG acknowledges the financial support from the Department of Science and Technology, India, through INSPIRE faculty award 
IFA17-PH197 at ARIES, Nainital.  
MFG acknowledges support from the National Science Foundation of China (grants 11873073 and U1531245). 
 This work makes extensive use of SDSS-I/II data. 
The SDSS Web Site is http://www.sdss.org/.

{

\appendix

\begin{table*}
\caption{ Estimates of optical measurements for the radio-loud sample. }
\begin{tabular}{cccccccccc} \hline
	RA        &Dec     &z     &FWHM (H$\alpha$)   &log L$_{H\alpha}$ &log M$_{\rm BH}$  &$\it \sigma_{\rm [SII]}$ 
&$\it \sigma_{\rm [OIII]}$ &log L$_{\rm [OIII]}$   &log $\nu$ L$_{\rm 1.4 GHz}$\\  
                  &   &   &($\rm km ~s^{-1}$) &$\rm erg ~s^{-1}$   &M$_{\odot}$     &$\rm km ~s^{-1}$   &$\rm km ~s^{-1}$    &$\rm erg ~s^{-1}$   &$\rm erg ~s^{-1}$ \\ \hline 
151.02 &  -0.55 &0.288 & 1896.99$\pm $ 149.80 &42.66$\pm$0.109  &7.44   & 69.41 $\pm$  2.65 & 112.34$\pm$ 34.80   &40.72 $\pm$1.16 &39.93      \\
172.59 &   0.97 &0.133 & 6734.06$\pm $  93.42 &42.31$\pm$0.005  &8.42   & 96.30 $\pm$  0.83 &  93.52$\pm$  2.38   &42.05 $\pm$0.04 &41.53      \\
205.31 &  -0.89 &0.237 & 2936.77$\pm $ 946.83 &43.45$\pm$0.262  &8.22   &333.71 $\pm$120.85 & 233.69$\pm$  3.95   &42.62 $\pm$0.02 &40.05      \\
207.11 &  -1.00 &0.234 & 4452.27$\pm $  95.62 &42.71$\pm$0.010  &8.24   &173.13 $\pm$  6.47 & 155.68$\pm$ 54.50   &41.44 $\pm$0.37 &39.27      \\
179.47 &  -3.26 &0.214 & 2677.75$\pm $1118.73 &43.18$\pm$0.185  &8.00   &333.16 $\pm$ 35.24 & 439.77$\pm$ 13.83   &42.37 $\pm$0.05 &39.72      \\
180.61 &  -1.49 &0.150 & 2435.38$\pm $ 245.82 &42.13$\pm$0.143  &7.43   &173.82 $\pm$ 18.62 & 150.46$\pm$ 98.74   &41.27 $\pm$0.68 &39.99      \\
357.98 &  -1.15 &0.174 & 4972.36$\pm $ 229.92 &43.61$\pm$0.039  &8.75   &159.90 $\pm$ 21.25 & 155.24$\pm$  6.78   &42.18 $\pm$0.05 &42.20      \\
142.16 &  60.42 &0.295 & 5476.07$\pm $ 662.55 &43.66$\pm$0.046  &8.87   &274.31 $\pm$ 76.39 & 199.70$\pm$ 39.69   &42.53 $\pm$0.02 &41.97      \\
166.41 &   2.05 &0.106 & 3774.39$\pm $  62.15 &42.83$\pm$0.003  &8.15   & 86.42 $\pm$  8.01 & 156.28$\pm$  5.08   &41.39 $\pm$0.05 &40.90      \\
168.86 &   1.57 &0.239 & 4584.40$\pm $ 348.21 &42.52$\pm$0.036  &8.18   &199.81 $\pm$  9.37 & 186.35$\pm$  6.33   &41.93 $\pm$0.04 &39.51      \\
185.05 &   2.06 &0.240 & 3999.25$\pm $ 214.98 &43.95$\pm$0.031  &8.73   &207.42 $\pm$ 41.59 & 139.84$\pm$  5.43   &42.60 $\pm$0.03 &41.91      \\
203.22 &   2.01 &0.216 & 3967.36$\pm $ 410.21 &42.52$\pm$0.020  &8.04   &135.23 $\pm$  7.45 & 123.49$\pm$  7.71   &41.56 $\pm$0.26 &41.80      \\
226.09 &   1.87 &0.183 & 2520.68$\pm $  76.36 &42.28$\pm$0.044  &7.52   &131.44 $\pm$  3.41 & 170.81$\pm$  3.67   &41.79 $\pm$0.03 &39.55      \\
117.74 &  35.51 &0.176 & 3181.87$\pm $  29.91 &42.17$\pm$0.006  &7.68   &139.94 $\pm$  0.78 & 126.95$\pm$  3.33   &40.67 $\pm$0.08 &39.54      \\
121.04 &  38.90 &0.211 & 2205.26$\pm $ 948.58 &43.29$\pm$0.135  &7.86   &196.03 $\pm$ 55.98 & 140.83$\pm$ 19.26   &42.26 $\pm$0.80 &39.64      \\
134.54 &  49.43 &0.174 & 2499.33$\pm $  29.95 &42.38$\pm$0.011  &7.57   &141.54 $\pm$  2.22 & 139.15$\pm$ 27.31   &41.19 $\pm$0.40 &39.10      \\
133.41 &   3.55 &0.208 & 1978.27$\pm $  93.18 &42.16$\pm$0.023  &7.26   & 91.24 $\pm$  1.91 &  55.67$\pm$  5.67   &40.14 $\pm$0.53 &39.81      \\
206.57 &  62.35 &0.116 & 5507.92$\pm $  45.45 &42.63$\pm$0.003  &8.39   &119.64 $\pm$  1.95 & 116.90$\pm$  7.99   &41.78 $\pm$0.01 &40.81      \\
229.00 &  57.40 &0.204 & 2928.37$\pm $ 183.20 &42.35$\pm$0.069  &7.70   &154.48 $\pm$  6.98 & 168.77$\pm$ 41.57   &41.05 $\pm$0.30 &39.76      \\
 17.79 & -10.28 &0.178 & 3794.59$\pm $1173.18 &43.10$\pm$0.116  &8.26   &379.93 $\pm$ 99.95 & 435.04$\pm$103.97   &41.19 $\pm$0.33 &40.05      \\
 35.06 &  -7.48 &0.213 & 5820.25$\pm $ 170.28 &42.88$\pm$0.020  &8.55   &132.50 $\pm$ 10.29 & 259.88$\pm$ 91.59   &41.41 $\pm$0.79 &39.22      \\
338.66 &  -9.36 &0.247 & 4003.20$\pm $  35.13 &42.88$\pm$0.006  &8.22   &142.37 $\pm$  1.70 & 156.84$\pm$  3.33   &42.27 $\pm$0.02 &39.47      \\
236.57 &  50.14 &0.286 & 4333.21$\pm $ 454.42 &42.37$\pm$0.035  &8.04   &180.10 $\pm$  4.14 & 154.36$\pm$  7.25   &42.03 $\pm$0.11 &40.93      \\
130.52 &  40.31 &0.152 & 7496.17$\pm $  87.48 &43.02$\pm$0.004  &8.85   &194.80 $\pm$  6.03 & 157.91$\pm$ 13.27   &41.39 $\pm$0.07 &40.25      \\
190.85 &   5.25 &0.165 & 3208.00$\pm $ 162.81 &42.91$\pm$0.042  &8.04   &411.24 $\pm$ 29.80 & 232.59$\pm$ 35.99   &41.49 $\pm$1.52 &40.15      \\
192.57 &   4.96 &0.233 & 3002.10$\pm $ 299.69 &42.77$\pm$0.035  &7.90   &260.00 $\pm$ 20.91 & 131.44$\pm$ 15.28   &41.12 $\pm$1.93 &39.50      \\
195.03 &   3.93 &0.184 &12193.50$\pm $ 211.03 &42.78$\pm$0.024  &9.17   &242.55 $\pm$  1.91 & 239.20$\pm$ 13.40   &42.48 $\pm$0.08 &40.41      \\
178.03 &  52.19 &0.289 & 5464.43$\pm $ 218.65 &43.08$\pm$0.016  &8.59   &217.52 $\pm$  4.89 & 198.58$\pm$  3.88   &42.53 $\pm$0.02 &39.78      \\
184.06 &  52.71 &0.270 & 2944.23$\pm $  88.25 &43.18$\pm$0.015  &8.09   &127.41 $\pm$  5.31 & 166.98$\pm$  4.05   &42.14 $\pm$0.03 &40.34      \\
202.14 &  -2.56 &0.184 & 4820.96$\pm $ 513.40 &41.97$\pm$0.024  &7.93   &229.98 $\pm$  1.61 & 236.44$\pm$ 43.01   &41.91 $\pm$0.15 &39.13      \\
227.59 &  -2.07 &0.268 & 4313.44$\pm $ 410.56 &42.87$\pm$0.045  &8.28   &105.24 $\pm$  5.78 & 138.23$\pm$  6.97   &41.95 $\pm$0.07 &39.46      \\
140.08 &  39.87 &0.223 & 4382.65$\pm $ 515.16 &42.50$\pm$0.045  &8.12   &183.22 $\pm$ 16.80 & 192.31$\pm$ 88.41   &41.21 $\pm$0.30 &40.25      \\
148.92 &  45.54 &0.259 & 4729.22$\pm $  77.03 &43.46$\pm$0.003  &8.65   &111.88 $\pm$  6.17 & 147.68$\pm$  5.85   &42.44 $\pm$0.04 &41.13      \\
201.08 &  58.82 &0.192 & 5456.72$\pm $ 239.92 &42.91$\pm$0.021  &8.51   &296.40 $\pm$ 16.70 & 280.81$\pm$ 12.91   &42.01 $\pm$0.06 &39.91      \\
166.82 &   8.08 &0.200 & 3278.49$\pm $  67.54 &43.12$\pm$0.014  &8.15   &142.96 $\pm$  3.83 & 168.24$\pm$ 58.53   &41.36 $\pm$0.76 &40.30      \\
185.81 &  54.15 &0.156 & 5307.53$\pm $ 172.18 &42.85$\pm$0.045  &8.45   &185.13 $\pm$  6.75 &  89.71$\pm$ 26.45   &42.06 $\pm$0.27 &41.52      \\
215.81 &  50.93 &0.275 & 7931.44$\pm $ 625.48 &43.27$\pm$0.005  &9.02   &190.51 $\pm$  2.31 & 113.64$\pm$  4.48   &42.48 $\pm$0.36 &41.61      \\
206.57 &  58.00 &0.163 & 3125.35$\pm $ 846.36 &42.46$\pm$0.230  &7.73   &302.31 $\pm$ 30.33 & 208.02$\pm$ 77.99   &41.38 $\pm$0.74 &39.20      \\
220.24 &  54.97 &0.231 & 2183.12$\pm $  73.85 &42.30$\pm$0.080  &7.41   &261.79 $\pm$  3.28 & 263.25$\pm$ 31.91   &41.30 $\pm$0.14 &39.44      \\
227.88 &  50.37 &0.221 & 2170.34$\pm $  60.70 &42.39$\pm$0.013  &7.45   & 77.50 $\pm$  1.23 &  59.67$\pm$ 15.97   &40.20 $\pm$1.23 &39.64      \\
229.78 &  52.10 &0.138 & 1793.59$\pm $ 155.40 &41.70$\pm$0.113  &6.94   &221.58 $\pm$  2.72 & 206.52$\pm$  7.75   &41.76 $\pm$0.04 &39.76      \\
236.32 &  46.64 &0.228 & 3846.55$\pm $ 119.95 &43.14$\pm$0.013  &8.31   &186.48 $\pm$  3.16 & 178.21$\pm$  9.38   &41.68 $\pm$0.07 &39.85      \\
248.81 &  36.45 &0.211 & 3205.99$\pm $  65.22 &42.52$\pm$0.012  &7.85   &141.03 $\pm$  2.17 & 144.21$\pm$ 35.76   &41.82 $\pm$0.49 &39.42      \\
144.09 &  39.36 &0.210 & 2501.88$\pm $ 432.51 &42.87$\pm$0.033  &7.84   &164.05 $\pm$ 14.46 & 186.05$\pm$  3.58   &42.28 $\pm$0.02 &40.66      \\
128.20 &  28.89 &0.226 & 3114.53$\pm $  45.71 &42.89$\pm$0.006  &7.99   &160.60 $\pm$  2.37 & 130.89$\pm$  3.21   &42.27 $\pm$0.02 &39.51      \\
144.26 &  36.26 &0.180 & 2211.36$\pm $   6.87 &42.62$\pm$0.006  &7.57   & 98.89 $\pm$  0.96 & 129.76$\pm$ 10.91   &41.36 $\pm$0.29 &39.57      \\
191.05 &  50.70 &0.215 & 2431.26$\pm $ 245.62 &42.45$\pm$0.082  &7.57   &220.90 $\pm$ 12.23 & 190.85$\pm$ 40.00   &41.34 $\pm$0.39 &39.56      \\
130.52 &   7.99 &0.134 & 8048.88$\pm $  93.64 &43.06$\pm$0.003  &8.93   &219.10 $\pm$  1.42 & 158.08$\pm$  2.24   &42.32 $\pm$0.01 &39.99      \\
148.74 &   9.50 &0.298 & 4267.31$\pm $  50.04 &43.28$\pm$0.004  &8.47   &111.94 $\pm$  1.42 &  96.08$\pm$  1.57   &42.78 $\pm$0.02 &42.18      \\
250.88 &  30.81 &0.184 &10984.10$\pm $ 105.18 &42.83$\pm$0.033  &9.10   &242.04 $\pm$  5.00 & 310.75$\pm$ 10.35   &41.78 $\pm$0.04 &40.90      \\
164.15 &  41.77 &0.202 & 6992.39$\pm $  26.43 &43.32$\pm$0.003  &8.93   &351.78 $\pm$  2.35 & 274.05$\pm$  2.27   &42.98 $\pm$0.00 &40.30      \\
168.73 &  42.54 &0.257 & 6853.98$\pm $ 192.04 &43.27$\pm$0.049  &8.88   &188.86 $\pm$  7.32 & 301.07$\pm$ 11.75   &42.22 $\pm$0.05 &39.69      \\
171.55 &  42.88 &0.157 & 4961.33$\pm $  75.34 &42.75$\pm$0.016  &8.35   &163.90 $\pm$  9.64 & 257.69$\pm$ 45.38   &41.82 $\pm$0.06 &38.97      \\
174.22 &  44.84 &0.116 & 3653.79$\pm $ 644.66 &42.01$\pm$0.045  &7.77   &186.31 $\pm$ 10.01 & 131.66$\pm$ 10.88   &41.27 $\pm$0.11 &39.19      \\
200.92 &  42.90 &0.177 & 6252.06$\pm $ 245.64 &42.29$\pm$0.016  &8.36   &190.67 $\pm$  1.80 & 174.82$\pm$ 27.63   &41.29 $\pm$0.44 &39.43      \\
243.00 &  31.18 &0.159 &10844.80$\pm $  79.25 &42.74$\pm$0.004  &9.04   &271.88 $\pm$  4.41 & 205.07$\pm$ 11.43   &41.67 $\pm$0.08 &39.82      \\
163.83 &  39.48 &0.253 & 5765.80$\pm $ 526.50 &43.11$\pm$0.028  &8.66   &347.63 $\pm$ 15.54 & 347.31$\pm$ 33.05   &42.38 $\pm$0.17 &39.88      \\
167.36 &  42.43 &0.232 & 4759.29$\pm $ 355.89 &42.67$\pm$0.056  &8.28   &271.44 $\pm$  6.77 & 222.17$\pm$ 30.61   &41.48 $\pm$0.59 &39.54      \\
175.20 &  46.37 &0.115 & 2671.59$\pm $  45.11 &42.90$\pm$0.008  &7.87   &132.20 $\pm$  2.09 & 142.20$\pm$ 18.84   &41.14 $\pm$0.15 &40.61      \\
178.13 &  42.47 &0.250 & 5704.49$\pm $ 106.42 &43.01$\pm$0.009  &8.60   &184.72 $\pm$  1.49 & 172.31$\pm$  1.59   &42.82 $\pm$0.01 &39.86      \\
179.37 &  43.30 &0.230 & 2760.57$\pm $  43.78 &42.86$\pm$0.006  &7.88   &130.27 $\pm$  1.77 & 136.50$\pm$  3.29   &42.47 $\pm$0.03 &41.70      \\

\hline

\end{tabular}  \\
\end{table*}

\begin{table*}
\addtocounter{figure}{-1}
\caption{ Table 1 continued... }
\begin{tabular}{cccccccccc} \hline

        RA        &Dec     &z     &FWHM (H$\alpha$)   &log L$_{H\alpha}$ &log M$_{\rm BH}$  &$\it \sigma_{\rm [SII]}$
&$\it \sigma_{\rm [OIII]}$ &log L$_{\rm [OIII]}$   &log $\nu$ L$_{\rm 1.4 GHz}$\\
                  &   &   &($\rm km ~s^{-1}$) &$\rm erg ~s^{-1}$   &M$_{\odot}$     &$\rm km ~s^{-1}$   &$\rm km ~s^{-1}$    &$\rm erg ~s^{-1}$   &$\rm erg ~s^{-1}$ \\ \hline

183.84 &  41.77 &0.196 & 5201.69$\pm $2154.36 &42.84$\pm$0.135  &8.42   &324.79 $\pm$ 29.94 & 148.68$\pm$ 17.44   &41.90 $\pm$0.22 &40.38      \\
358.22 &   0.64 &0.273 & 3613.83$\pm $ 200.19 &42.79$\pm$0.022  &8.10   &170.06 $\pm$ 10.11 & 209.14$\pm$11.48    &41.50 $\pm$0.61 &39.84      \\
254.92 &  18.58 &0.171 & 7595.62$\pm $ 273.42 &42.96$\pm$0.027  &8.83   &194.20 $\pm$  2.73 & 230.19$\pm$  1.39   &42.46 $\pm$0.01 &39.42      \\
178.00 &  12.36 &0.170 & 4507.37$\pm $  62.93 &42.72$\pm$0.013  &8.25   &167.42 $\pm$  6.08 & 159.24$\pm$  3.86   &41.48 $\pm$0.28 &39.43      \\
183.21 &   7.00 &0.209 & 3127.04$\pm $  49.86 &42.70$\pm$0.012  &7.92   &211.15 $\pm$  5.01 & 168.15$\pm$  7.28   &41.89 $\pm$0.05 &39.54      \\
184.94 &   8.35 &0.229 & 2801.76$\pm $  45.42 &42.78$\pm$0.011  &7.86   & 91.00 $\pm$  1.51 &  94.77$\pm$ 10.34   &40.75 $\pm$0.34 &39.78      \\
243.67 &  23.21 &0.294 & 2960.76$\pm $ 102.67 &43.32$\pm$0.018  &8.16   &215.12 $\pm$  7.51 & 169.77$\pm$  3.10   &42.81 $\pm$0.02 &40.16      \\
248.75 &  20.83 &0.128 & 2333.71$\pm $ 523.29 &42.59$\pm$0.052  &7.62   &191.57 $\pm$ 19.27 & 459.60$\pm$ 66.22   &40.99 $\pm$0.98 &40.09      \\
200.28 &  50.78 &0.233 & 1677.28$\pm $  55.62 &42.74$\pm$0.029  &7.37   &172.10 $\pm$  2.82 & 194.90$\pm$ 16.78   &41.62 $\pm$0.14 &39.88      \\
213.99 &  49.89 &0.185 & 2204.87$\pm $  33.54 &42.33$\pm$0.011  &7.43   & 97.30 $\pm$  0.80 & 112.34$\pm$  6.30   &41.60 $\pm$0.20 &40.69      \\
222.52 &  46.59 &0.292 & 3530.41$\pm $ 820.92 &43.39$\pm$0.118  &8.41   &339.63 $\pm$ 35.81 & 260.95$\pm$ 28.86   &42.89 $\pm$0.22 &40.28      \\
255.49 &  22.45 &0.197 & 5475.30$\pm $  61.19 &42.70$\pm$0.004  &8.42   &121.31 $\pm$  1.28 & 115.14$\pm$  8.29   &41.97 $\pm$0.35 &40.58      \\
254.24 &  25.42 &0.267 & 2338.84$\pm $ 477.70 &41.88$\pm$0.292  &7.27   &279.95 $\pm$  7.64 & 141.95$\pm$ 18.27   &40.95 $\pm$0.70 &41.60      \\
243.07 &   7.53 &0.207 & 3661.10$\pm $1053.59 &43.22$\pm$0.031  &8.27   &170.21 $\pm$ 26.37 & 119.25$\pm$  4.11   &42.07 $\pm$0.02 &39.87      \\
117.28 &  45.18 &0.192 & 3541.23$\pm $ 131.96 &43.46$\pm$0.009  &8.38   &136.84 $\pm$  8.12 & 141.59$\pm$  1.33   &42.60 $\pm$0.01 &41.20      \\
201.29 &   6.19 &0.183 & 3803.38$\pm $ 337.43 &42.69$\pm$0.025  &8.08   &210.66 $\pm$  8.30 & 192.69$\pm$  3.16   &42.35 $\pm$0.02 &39.24      \\
223.64 &   8.06 &0.130 & 2580.38$\pm $ 162.26 &42.88$\pm$0.057  &7.83   &277.83 $\pm$ 12.67 & 136.34$\pm$  8.04   &41.67 $\pm$0.06 &39.86      \\
228.72 &   5.61 &0.174 & 3641.34$\pm $ 203.90 &42.28$\pm$0.040  &7.86   &194.34 $\pm$  4.44 & 179.32$\pm$  3.83   &41.80 $\pm$0.02 &39.47      \\
233.12 &   4.90 &0.218 & 7351.62$\pm $1808.00 &43.07$\pm$0.043  &8.88   &220.34 $\pm$ 13.30 & 176.10$\pm$  3.68   &42.37 $\pm$0.02 &40.33      \\
150.53 &  34.90 &0.205 & 2459.14$\pm $ 577.28 &42.67$\pm$0.081  &7.73   &229.33 $\pm$ 19.93 & 312.36$\pm$ 26.82   &42.32 $\pm$0.21 &39.91      \\
157.75 &  31.05 &0.178 & 5661.81$\pm $  36.94 &43.40$\pm$0.013  &8.77   &122.89 $\pm$  4.26 & 104.52$\pm$  1.46   &42.32 $\pm$0.02 &41.39      \\
186.60 &  32.74 &0.243 & 3908.93$\pm $ 234.40 &43.05$\pm$0.026  &8.28   &214.81 $\pm$ 11.08 & 205.98$\pm$  5.45   &42.20 $\pm$0.03 &39.79      \\
200.08 &  39.12 &0.236 & 3251.36$\pm $  88.88 &42.44$\pm$0.024  &7.83   &177.99 $\pm$  2.26 & 146.29$\pm$  1.95   &42.40 $\pm$0.01 &39.58      \\
191.08 &  40.86 &0.249 & 3047.16$\pm $ 109.65 &42.45$\pm$0.025  &7.77   &209.44 $\pm$  3.94 & 172.59$\pm$  7.49   &41.80 $\pm$0.05 &41.93      \\
204.10 &  39.29 &0.179 & 4138.06$\pm $  90.02 &43.15$\pm$0.012  &8.38   &161.84 $\pm$  2.93 & 165.00$\pm$  5.18   &42.12 $\pm$0.03 &39.90      \\
186.96 &  32.25 &0.137 & 1913.74$\pm $ 795.33 &42.77$\pm$0.180  &7.56   &192.33 $\pm$ 44.35 & 123.74$\pm$  2.42   &42.01 $\pm$0.01 &39.62      \\
171.16 &  37.38 &0.227 & 4667.96$\pm $ 257.15 &42.57$\pm$0.015  &8.21   &107.40 $\pm$  4.19 & 119.90$\pm$  1.97   &42.07 $\pm$0.02 &41.48      \\
206.91 &  30.21 &0.118 & 4949.12$\pm $ 252.29 &43.09$\pm$0.010  &8.50   & 93.60 $\pm$  5.88 &  76.90$\pm$  2.37   &41.99 $\pm$0.02 &39.87      \\
181.82 &  36.80 &0.250 & 3248.87$\pm $ 288.30 &42.99$\pm$0.044  &8.09   &179.99 $\pm$ 19.48 & 161.51$\pm$  5.30   &42.19 $\pm$0.02 &39.38      \\
184.64 &  35.38 &0.240 & 3908.28$\pm $ 178.56 &43.08$\pm$0.008  &8.29   &124.41 $\pm$  6.21 & 144.57$\pm$ 47.76   &42.07 $\pm$0.03 &39.67      \\
166.77 &  32.11 &0.242 & 3099.58$\pm $1961.06 &43.47$\pm$0.065  &8.08   &280.42 $\pm$ 46.70 & 311.29$\pm$  3.85   &43.03 $\pm$0.01 &39.79      \\
211.75 &  28.45 &0.077 & 7900.27$\pm $1155.93 &43.11$\pm$0.056  &8.94   &240.81 $\pm$ 32.95 & 218.24$\pm$  8.52   &41.62 $\pm$0.05 &41.21      \\
215.63 &  29.87 &0.113 & 2309.35$\pm $ 485.05 &42.42$\pm$0.097  &7.50   &245.31 $\pm$  8.19 & 228.69$\pm$  7.39   &41.89 $\pm$0.03 &39.85      \\
217.38 &  25.09 &0.145 & 2441.87$\pm $  58.35 &42.42$\pm$0.017  &7.56   &221.38 $\pm$  6.35 & 211.32$\pm$ 99.78   &41.48 $\pm$0.65 &38.94      \\
219.26 &  26.67 &0.218 & 7347.50$\pm $ 180.28 &42.93$\pm$0.006  &8.79   &188.12 $\pm$  2.76 & 151.28$\pm$ 74.02   &41.58 $\pm$0.40 &40.16      \\
161.27 &   8.73 &0.125 & 3369.40$\pm $ 137.05 &43.20$\pm$0.016  &8.23   &160.83 $\pm$ 16.18 & 140.12$\pm$  2.23   &42.25 $\pm$0.01 &39.74      \\
232.40 &  26.04 &0.251 & 2319.59$\pm $  83.45 &42.62$\pm$0.014  &7.61   &113.04 $\pm$  2.72 & 114.16$\pm$  3.29   &42.21 $\pm$0.03 &40.45      \\
236.14 &  20.44 &0.267 & 2113.28$\pm $ 199.94 &43.07$\pm$0.037  &7.72   &138.08 $\pm$  2.92 & 141.11$\pm$  1.91   &42.63 $\pm$0.01 &39.53      \\
189.55 &  27.22 &0.236 & 4869.22$\pm $ 142.86 &42.28$\pm$0.020  &8.12   &186.50 $\pm$  2.43 & 166.97$\pm$ 17.01   &41.45 $\pm$0.08 &39.38      \\
190.05 &  31.05 &0.236 & 3826.42$\pm $ 137.53 &42.41$\pm$0.027  &7.95   &226.42 $\pm$  5.86 & 210.33$\pm$24.23    &41.40 $\pm$0.87 &39.42      \\
139.70 &  21.29 &0.149 & 4415.76$\pm $3647.61 &42.78$\pm$0.521  &8.17   &310.10 $\pm$108.00 & 167.76$\pm$ 41.71   &41.78 $\pm$0.24 &39.66      \\
143.45 &  21.24 &0.172 & 3811.31$\pm $  38.58 &42.77$\pm$0.005  &8.13   &153.54 $\pm$  2.35 & 162.99$\pm$  7.80   &41.74 $\pm$0.06 &39.47      \\
146.80 &  22.48 &0.181 & 2968.77$\pm $ 253.12 &42.64$\pm$0.073  &7.85   &269.31 $\pm$ 12.28 & 157.56$\pm$  3.91   &41.73 $\pm$0.03 &39.43      \\
148.53 &  21.38 &0.295 & 8199.97$\pm $ 122.81 &43.52$\pm$0.031  &9.16   &131.21 $\pm$  8.73 & 125.62$\pm$  2.55   &42.54 $\pm$0.02 &40.97      \\
121.43 &  11.51 &0.199 & 3475.58$\pm $ 350.86 &42.75$\pm$0.058  &8.02   &198.92 $\pm$ 18.31 & 103.64$\pm$  1.99   &41.85 $\pm$0.02 &39.58      \\
120.51 &  10.33 &0.205 & 2414.66$\pm $  13.67 &42.87$\pm$0.006  &7.76   &174.88 $\pm$  1.56 & 119.57$\pm$  2.45   &42.27 $\pm$0.02 &39.90      \\
134.70 &  15.98 &0.155 & 7685.88$\pm $1757.58 &42.38$\pm$0.031  &8.57   &160.92 $\pm$  8.38 & 147.53$\pm$  2.30   &41.96 $\pm$0.02 &39.02      \\
159.54 &  23.53 &0.226 & 3552.99$\pm $ 109.42 &42.98$\pm$0.022  &8.16   &271.42 $\pm$  3.48 & 231.34$\pm$  4.60   &42.67 $\pm$0.02 &39.97      \\
164.38 &  18.36 &0.232 & 7521.62$\pm $ 136.54 &42.95$\pm$0.008  &8.82   &213.94 $\pm$  2.96 & 191.78$\pm$  8.84   &42.03 $\pm$0.06 &39.48      \\
166.15 &  21.41 &0.188 & 3732.08$\pm $  90.22 &43.19$\pm$0.011  &8.31   &135.39 $\pm$  8.11 & 124.23$\pm$ 28.80   &41.89 $\pm$0.03 &40.86      \\
249.44 &  11.83 &0.147 & 3851.17$\pm $  45.27 &42.95$\pm$0.004  &8.22   & 99.56 $\pm$  7.49 &  83.31$\pm$  7.14   &41.49 $\pm$0.11 &39.96      \\
204.96 &  15.99 &0.276 & 9039.28$\pm $ 389.90 &43.19$\pm$0.011  &9.10   &233.19 $\pm$  3.10 & 215.69$\pm$ 31.02   &42.69 $\pm$0.03 &39.83      \\
189.94 &  19.91 &0.239 &10790.60$\pm $ 117.11 &43.26$\pm$0.021  &9.29   &195.55 $\pm$  1.87 & 208.22$\pm$  0.84   &42.90 $\pm$0.00 &39.71      \\
196.09 &  20.92 &0.283 & 4339.29$\pm $ 431.94 &43.00$\pm$0.044  &8.34   &187.32 $\pm$ 21.40 & 153.58$\pm$  5.06   &42.23 $\pm$0.03 &39.76      \\
213.68 &  16.97 &0.238 & 3933.07$\pm $  94.81 &43.14$\pm$0.003  &8.33   &122.84 $\pm$  5.18 & 179.03$\pm$  1.93   &42.34 $\pm$0.01 &39.48      \\
212.59 &  20.25 &0.202 & 4971.48$\pm $ 570.89 &42.88$\pm$0.032  &8.41   &261.92 $\pm$ 16.91 & 211.44$\pm$ 13.11   &41.92 $\pm$0.10 &39.56      \\
235.03 &  14.19 &0.119 & 4402.15$\pm $ 155.96 &42.85$\pm$0.014  &8.29   &191.53 $\pm$ 38.91 & 342.36$\pm$159.19   &41.44 $\pm$0.95 &40.08      \\
212.67 &  22.56 &0.172 & 2777.57$\pm $ 309.91 &42.88$\pm$0.053  &7.89   &309.48 $\pm$ 96.72 & 261.44$\pm$ 11.28   &41.79 $\pm$0.05 &39.59      \\
215.97 &  22.80 &0.281 & 9681.19$\pm $ 635.03 &43.12$\pm$0.044  &9.14   &229.36 $\pm$ 19.00 & 210.18$\pm$  3.17   &42.45 $\pm$0.01 &39.84      \\
227.31 &  17.95 &0.171 & 1978.45$\pm $  64.05 &43.11$\pm$0.015  &7.69   &140.99 $\pm$  2.93 & 145.48$\pm$  2.56   &42.13 $\pm$0.02 &40.09      \\
186.92 &   4.33 &0.180 & 3385.01$\pm $ 439.01 &42.85$\pm$0.061  &8.05   &220.38 $\pm$ 10.78 & 171.38$\pm$  1.62   &42.43 $\pm$0.01 &39.91      \\
236.40 &   1.83 &0.250 & 7982.62$\pm $2857.48 &42.41$\pm$0.023  &8.63   &187.39 $\pm$  7.43 & 160.96$\pm$  5.60   &42.03 $\pm$0.05 &42.30      \\
\hline
\end{tabular}  \\
\end{table*}

\begin{table*}
\caption{ Table 1 continued... }
\begin{tabular}{cccccccccc} \hline

        RA        &Dec     &z     &FWHM (H$\alpha$)   &log L$_{H\alpha}$ &log M$_{\rm BH}$  &$\it \sigma_{\rm [SII]}$
&$\it \sigma_{\rm [OIII]}$ &log L$_{\rm [OIII]}$   &log $\nu$ L$_{\rm 1.4 GHz}$\\
                  &   &   &($\rm km ~s^{-1}$) &$\rm erg ~s^{-1}$   &M$_{\odot}$     &$\rm km ~s^{-1}$   &$\rm km ~s^{-1}$    &$\rm erg ~s^{-1}$   &$\rm erg ~s^{-1}$ \\ \hline
141.65 &   7.41 &0.190 & 1073.08$\pm $  11.08 &41.37$\pm$0.255   &6.38  &   & 123.98$\pm$ 49.54   &41.17 $\pm$0.86 &39.21      \\
220.67 &   5.41 &0.117 & 3660.99$\pm $   700.00 &43.24$\pm$0.28  &8.31  &  & 177.56$\pm$  6.61   &41.27 $\pm$0.04 &39.36      \\
203.71 &  1.04 &0.246 & 2363.23$\pm$     973.95 &42.32$\pm$0.359 &7.31  &  & 190.51$\pm$  110.43    &40.59 $\pm$0.931          &39.58         \\
225.11 & -0.91 &0.281 & 4457.04$\pm$      94.13 &42.89$\pm$0.084 &8.19  &  &  69.14$\pm$   68.89    &41.55 $\pm$1.511          &39.64         \\
225.11 & -0.91 &0.281 & 5182.03$\pm$     229.97 &42.32$\pm$0.094 &8.41  &  & 167.43$\pm$   21.13    &41.42 $\pm$0.236          &39.64         \\
198.36 & -2.54 &0.174 & 2447.77$\pm$     244.79 &42.27$\pm$0.273 &8.39  &  & 126.84$\pm$  153.42    &40.78 $\pm$1.256          &39.25         \\
198.36 & -2.54 &0.174 & 6523.64$\pm$     273.37 &42.32$\pm$0.019 &8.35  &  & 299.27$\pm$  106.66    &40.51 $\pm$0.802          &39.25         \\
236.11 &  0.16 &0.281 & 6489.82$\pm$     377.19 &42.71$\pm$0.064 &9.21  &  & 175.07$\pm$   25.48    &41.34 $\pm$0.221          &40.35         \\
238.03 & -0.89 &0.298 &13726.30$\pm$    1656.22 &42.71$\pm$0.512 &7.88  &  & 134.60$\pm$    9.98    &41.59 $\pm$0.118          &41.56         \\
120.84 & 43.55 &0.275 & 4927.76$\pm$     209.31 &44.06$\pm$2.191 &8.15  &  & 182.49$\pm$  177.51    &40.97 $\pm$1.363          &40.23         \\
120.84 & 43.55 &0.275 & 3677.58$\pm$     379.53 &42.89$\pm$0.651 &8.17  &  & 127.31$\pm$  126.69    &40.81 $\pm$1.494          &40.22         \\
120.38 & 47.60 &0.157 & 3657.70$\pm$     167.89 &42.39$\pm$0.073 &9.00  &  & 134.21$\pm$   11.29    &41.61 $\pm$0.092          &40.74         \\
146.52 &  1.66 &0.220 & 7109.96$\pm$     546.24 &43.78$\pm$1.907 &8.29  &  & 226.84$\pm$    7.19    &42.10 $\pm$0.037          &40.14         \\
153.76 &  2.52 &0.218 & 5482.12$\pm$     733.03 &43.36$\pm$1.188 &7.82  &  &  97.22$\pm$  113.59    &40.28 $\pm$2.001          &40.22         \\
139.15 & 54.24 &0.284 & 2500.72$\pm$      27.76 &42.86$\pm$0.085 &8.10  &  & 166.23$\pm$   21.21    &41.54 $\pm$0.229          &40.56         \\
143.00 & 55.56 &0.265 & 3715.87$\pm$     848.14 &42.69$\pm$0.569 &8.51  &  & 120.79$\pm$  162.76    &40.76 $\pm$1.869          &40.27         \\
233.87 & 53.36 &0.291 & 5505.63$\pm$      42.66 &43.52$\pm$0.125 &7.93  &  & 102.70$\pm$    8.39    &41.70 $\pm$0.122          &40.49         \\
233.87 & 53.36 &0.291 & 2971.85$\pm$     178.96 &42.70$\pm$0.324 &7.93  &  & 120.59$\pm$    8.10    &41.76 $\pm$0.105          &40.49         \\
242.65 & 48.01 &0.247 & 2708.70$\pm$     189.00 &42.84$\pm$0.229 &8.42  &  & 146.34$\pm$    3.89    &42.08 $\pm$0.035          &40.22         \\
246.98 & 46.71 &0.214 & 2728.38$\pm$      46.74 &42.68$\pm$0.038 &8.55  &  & 338.39$\pm$  194.00    &40.80 $\pm$1.054          &39.48         \\
319.72 & -7.54 &0.260 & 5635.74$\pm$     199.06 &42.71$\pm$0.068 &7.27  &  & 141.30$\pm$    3.79    &42.00 $\pm$0.043          &41.39         \\
342.07 &-10.26 &0.291 & 5797.01$\pm$     230.31 &43.19$\pm$0.333 &8.80  &  & 138.75$\pm$   31.41    &42.18 $\pm$0.266          &40.82         \\
128.47 & 42.40 &0.249 & 4011.38$\pm$      47.13 &42.88$\pm$0.045 &6.90  &  & 106.75$\pm$  135.13    &40.65 $\pm$1.958          &41.75         \\
178.23 & 61.10 &0.292 & 5052.79$\pm$     354.80 &43.67$\pm$1.252 &7.57  &  & 146.87$\pm$  118.43    &41.01 $\pm$1.245          &39.82         \\
204.35 & 60.09 &0.234 & 1733.32$\pm$     519.50 &41.60$\pm$0.044 &8.62  &  & 223.94$\pm$  115.18    &41.61 $\pm$0.579          &39.96         \\
253.22 & 36.02 &0.281 & 3660.55$\pm$     400.00 &44.08$\pm$0.032 &7.96  &  & 111.86$\pm$   13.40    &41.29 $\pm$0.131          &39.83         \\
137.89 & 44.38 &0.298 & 3446.91$\pm$     141.00 &42.66$\pm$0.114 &8.28  &  & 165.23$\pm$   33.82    &41.69 $\pm$0.266          &42.17         \\
171.29 &  3.67 &0.261 & 7487.05$\pm$      72.66 &43.55$\pm$0.146 &7.70  &  & 386.03$\pm$  123.21    &41.75 $\pm$1.151          &40.63         \\
173.34 &  4.55 &0.248 & 4018.21$\pm$     398.40 &42.81$\pm$0.367 &7.50  &  &  90.54$\pm$  101.59    &41.39 $\pm$3.342          &40.06         \\
190.66 &  4.94 &0.233 & 3511.70$\pm$    1148.24 &41.92$\pm$0.183 &8.38  &  & 134.85$\pm$  116.03    &41.33 $\pm$0.998          &40.53         \\
188.63 & 51.94 &0.296 & 5464.27$\pm$     181.33 &43.11$\pm$0.176 &8.19  &  & 160.81$\pm$   62.83    &41.27 $\pm$0.473          &39.77         \\
113.84 & 26.80 &0.240 & 2968.18$\pm$      65.42 &43.06$\pm$0.147 &7.74  &  & 143.89$\pm$   34.31    &41.36 $\pm$0.319          &39.41         \\
144.30 & 50.15 &0.275 & 3660.53$\pm$      59.63 &44.06$\pm$0.572 &7.63  &  & 176.60$\pm$  102.52    &41.31 $\pm$0.721          &41.69         \\
151.04 & 52.51 &0.299 & 4510.00$\pm$     267.55 &42.60$\pm$0.125 &7.49  &  & 195.82$\pm$  146.64    &41.12 $\pm$1.258          &40.36         \\
151.04 & 52.51 &0.299 & 2700.81$\pm$      48.82 &42.82$\pm$0.115 &7.31  &  & 192.03$\pm$  134.08    &41.04 $\pm$0.833          &40.36         \\
164.04 & 55.27 &0.256 & 2812.26$\pm$     128.32 &42.22$\pm$0.060 &7.84  &  & 233.88$\pm$  177.65    &41.07 $\pm$1.315          &39.80         \\
203.64 & -1.64 &0.292 & 2061.83$\pm$     101.21 &42.18$\pm$0.058 &9.44  &  & 139.36$\pm$  112.95    &41.04 $\pm$0.840          &39.75         \\
117.33 & 23.68 &0.296 & 4696.50$\pm$     400.35 &42.43$\pm$0.067 &8.78  &  & 195.62$\pm$   10.25    &42.15 $\pm$0.083          &39.70         \\
117.33 & 23.68 &0.296 &15829.00$\pm$     355.41 &42.86$\pm$0.852 &8.90  &  & 206.75$\pm$    7.31    &42.24 $\pm$0.054          &39.70         \\
140.72 & 39.95 &0.289 & 7035.06$\pm$     623.10 &42.97$\pm$0.198 &7.61  &  & 106.70$\pm$    5.97    &41.25 $\pm$0.136          &40.21         \\
256.11 & 33.53 &0.290 & 2723.53$\pm$     238.14 &42.31$\pm$0.053 &8.14  &  & 232.87$\pm$  185.58    &41.68 $\pm$1.199          &41.01         \\
162.21 &  6.44 &0.275 & 4730.63$\pm$      79.29 &43.52$\pm$0.162 &8.02  &  & 318.91$\pm$  140.77    &41.65 $\pm$0.661          &39.57         \\
218.79 & 49.81 &0.166 & 3281.40$\pm$      64.42 &42.94$\pm$0.149 &8.45  &  & 180.19$\pm$   14.44    &41.32 $\pm$0.084          &39.46         \\
119.18 & 31.05 &0.271 & 5271.85$\pm$     203.23 &43.39$\pm$1.159 &8.27  &  & 172.51$\pm$  173.87    &39.91 $\pm$1.168          &40.80         \\
212.31 & 56.94 &0.238 & 6021.94$\pm$     367.75 &42.94$\pm$0.199 &7.74  &  & 273.66$\pm$  120.39    &41.59 $\pm$2.000          &39.77         \\
243.73 & 41.52 &0.196 & 2179.98$\pm$      52.08 &42.27$\pm$0.023 &7.73  &  & 260.76$\pm$  198.83    &40.77 $\pm$1.437          &39.84         \\
247.25 & 40.13 &0.272 & 1764.91$\pm$     140.03 &42.37$\pm$0.300 &7.73  &  & 127.64$\pm$   79.82    &42.06 $\pm$0.653          &40.53         \\
117.33 & 23.68 &0.296 & 2100.86$\pm$     901.66 &43.14$\pm$1.473 &8.87  &  & 194.55$\pm$    8.59    &42.13 $\pm$0.069          &39.70         \\
173.70 & 10.13 &0.262 & 1099.83$\pm$     219.36 &41.73$\pm$0.208 &8.84  &  & 115.27$\pm$  108.36    &41.67 $\pm$2.604          &40.13         \\
159.75 & 42.46 &0.220 & 8035.94$\pm$      88.11 &43.54$\pm$0.103 &8.15  &  & 299.35$\pm$   86.40    &41.34 $\pm$0.541          &39.64         \\
247.85 & 25.52 &0.270 & 6851.46$\pm$     131.13 &43.32$\pm$0.926 &8.76  &  & 181.66$\pm$   42.85    &41.78 $\pm$0.264          &39.87         \\
163.09 & 45.72 &0.240 & 3835.54$\pm$     574.97 &42.70$\pm$0.225 &8.81  &  & 177.69$\pm$   22.79    &41.83 $\pm$0.386          &41.38         \\
177.50 & 41.20 &0.250 & 8445.26$\pm$      90.93 &42.75$\pm$0.071 &7.74  &  & 155.58$\pm$    6.06    &42.04 $\pm$0.055          &41.39         \\
182.81 & 46.79 &0.294 & 2677.82$\pm$      35.81 &43.82$\pm$0.592 &8.47  &  & 261.12$\pm$   11.20    &41.84 $\pm$0.051          &39.87         \\
184.04 & 41.99 &0.242 & 2907.18$\pm$     455.29 &42.44$\pm$0.094 &8.88  &  & 152.70$\pm$    2.85    &42.20 $\pm$0.024          &40.89         \\
193.77 & 43.07 &0.300 & 2766.08$\pm$      44.02 &43.13$\pm$0.101 &8.15  &  & 109.18$\pm$   45.62    &41.64 $\pm$0.502          &39.64         \\
\hline

\end{tabular}  \\
\end{table*}

\begin{table*}
\caption{ Table 1 continued... }
\begin{tabular}{cccccccccc} \hline

        RA        &Dec     &z     &FWHM (H$\alpha$)   &log L$_{H\alpha}$ &log M$_{\rm BH}$  &$\it \sigma_{\rm [SII]}$
&$\it \sigma_{\rm [OIII]}$ &log L$_{\rm [OIII]}$   &log $\nu$ L$_{\rm 1.4 GHz}$\\
                  &   &   &($\rm km ~s^{-1}$) &$\rm erg ~s^{-1}$   &M$_{\odot}$     &$\rm km ~s^{-1}$   &$\rm km ~s^{-1}$    &$\rm erg ~s^{-1}$   &$\rm erg ~s^{-1}$ \\ \hline

237.58 & 30.63 &0.262 & 5108.08$\pm$    2020.86 &43.13$\pm$1.819 &8.67  &  & 227.89$\pm$  149.85    &41.59 $\pm$2.218          &39.60         \\
173.61 & 12.87 &0.295 & 4698.58$\pm$    2006.46 &42.41$\pm$0.251 &7.54  &  &  99.51$\pm$   56.28    &41.09 $\pm$1.479          &40.14         \\
219.63 & 35.65 &0.262 & 2437.36$\pm$     635.32 &42.26$\pm$0.292 &8.66  &  & 165.03$\pm$  113.72    &41.10 $\pm$1.252          &39.97         \\
223.50 &  9.26 &0.279 & 2206.26$\pm$      25.77 &42.73$\pm$0.059 &7.81  &  &  86.20$\pm$    3.89    &41.81 $\pm$0.047          &41.59         \\
113.77 & 43.20 &0.262 & 5488.50$\pm$      83.92 &42.98$\pm$0.065 &8.18  &  & 160.68$\pm$  192.71    &40.69 $\pm$1.883          &40.39         \\
151.86 & 12.82 &0.241 & 2610.41$\pm$      37.82 &42.68$\pm$0.048 &9.06  &  & 126.51$\pm$   10.49    &42.13 $\pm$0.093          &41.91         \\
156.85 & 12.32 &0.231 & 3541.37$\pm$     834.27 &43.33$\pm$0.556 &8.08  &  & 109.23$\pm$    5.80    &41.54 $\pm$0.066          &40.19         \\
134.14 & 59.96 &0.281 & 3334.10$\pm$     531.38 &43.22$\pm$1.115 &9.45  &  & 444.60$\pm$  108.70    &41.47 $\pm$0.619          &40.90         \\
194.41 &  8.16 &0.272 & 3541.13$\pm$     115.77 &43.80$\pm$0.597 &8.51  &  & 271.94$\pm$   67.19    &41.07 $\pm$0.308          &40.27         \\
203.76 &  9.72 &0.191 & 2839.47$\pm$      37.68 &43.06$\pm$0.133 &7.61  &  & 270.54$\pm$   31.44    &41.00 $\pm$0.153          &39.53         \\
224.15 & 31.88 &0.218 & 2595.63$\pm$     175.80 &43.37$\pm$1.531 &8.32  &  & 224.09$\pm$   87.05    &41.26 $\pm$0.435          &39.46         \\
238.42 & 24.30 &0.259 & 3660.99$\pm$       0.00 &44.00$\pm$0.000 &8.95  &  & 231.85$\pm$    8.37    &42.11 $\pm$0.046          &39.49         \\
129.26 & 25.14 &0.275 & 3656.13$\pm$     190.30 &42.73$\pm$0.239 &7.75  &  & 129.89$\pm$  100.86    &40.93 $\pm$0.932          &39.87         \\
142.91 & 32.07 &0.226 & 7570.80$\pm$    1776.35 &43.10$\pm$0.512 &8.73  &  & 474.31$\pm$    6.22    &41.51 $\pm$0.260          &40.25         \\
146.69 & 32.70 &0.236 & 5829.62$\pm$    1166.90 &42.45$\pm$1.068 &8.07  &  & 116.15$\pm$    1.41    &42.78 $\pm$0.015          &39.79         \\
177.50 & 41.20 &0.250 & 7450.61$\pm$    1782.11 &42.88$\pm$0.343 &7.75  &  & 158.23$\pm$    4.21    &42.03 $\pm$0.048          &41.39         \\
198.07 & 35.26 &0.183 & 3447.06$\pm$     347.37 &43.49$\pm$2.486 &8.53  &  & 216.07$\pm$    7.74    &42.33 $\pm$0.047          &40.73         \\
192.06 & 36.41 &0.207 & 3258.97$\pm$      74.59 &42.31$\pm$0.038 &8.61  &  & 110.24$\pm$    4.96    &41.63 $\pm$0.049          &40.27         \\
206.65 & 31.36 &0.246 & 3051.96$\pm$      96.30 &42.43$\pm$0.052 &7.26  &  & 163.81$\pm$   36.73    &41.26 $\pm$0.299          &40.59         \\
174.16 & 37.45 &0.265 & 4149.44$\pm$      86.38 &43.53$\pm$0.419 &8.78  &  & 172.83$\pm$    3.98    &42.15 $\pm$0.042          &40.76         \\
223.38 & 26.83 &0.279 & 2501.05$\pm$     844.41 &43.61$\pm$1.245 &9.38  &  & 198.17$\pm$    9.06    &42.17 $\pm$0.072          &40.63         \\
225.31 & 23.49 &0.258 & 2293.43$\pm$     537.89 &43.20$\pm$1.202 &7.86  &  & 167.47$\pm$  206.46    &41.63 $\pm$4.196          &40.00         \\
231.99 & 22.55 &0.254 & 2443.97$\pm$      57.27 &42.96$\pm$0.164 &8.12  &  & 134.12$\pm$    2.70    &42.22 $\pm$0.030          &41.91         \\
236.93 & 20.87 &0.264 & 3400.53$\pm$     159.94 &43.92$\pm$1.635 &8.98  &  & 208.13$\pm$   11.25    &42.82 $\pm$0.087          &42.64         \\
197.08 & 27.72 &0.237 & 2319.59$\pm$      75.30 &42.56$\pm$0.069 &7.74  &  &  88.94$\pm$    1.65    &41.83 $\pm$0.024          &39.43         \\
199.78 & 27.31 &0.284 & 2072.95$\pm$     186.91 &43.13$\pm$0.375 &8.44  &  & 117.55$\pm$   93.78    &41.18 $\pm$1.014          &39.63         \\
126.45 & 17.56 &0.223 & 5016.67$\pm$      79.19 &43.90$\pm$1.020 &8.36  &  & 280.42$\pm$  196.02    &40.85 $\pm$1.832          &40.15         \\
127.92 & 19.93 &0.211 & 4898.23$\pm$     131.79 &42.17$\pm$0.022 &8.14  &  & 218.34$\pm$  164.27    &41.16 $\pm$0.951          &39.90         \\
130.95 & 20.63 &0.227 & 3804.07$\pm$     142.29 &42.37$\pm$0.064 &9.37  &  & 427.36$\pm$   23.52    &41.69 $\pm$0.112          &39.70         \\
154.88 & 26.45 &0.250 & 4875.18$\pm$    1600.17 &42.49$\pm$0.170 &8.94  &  & 186.50$\pm$    9.37    &42.03 $\pm$0.063          &39.43         \\
141.79 & 16.47 &0.212 & 3917.86$\pm$    3511.35 &43.15$\pm$6.667 &8.08  &  & 137.13$\pm$   13.46    &41.36 $\pm$0.178          &39.45         \\
139.87 & 14.53 &0.207 & 8188.07$\pm$     121.51 &43.17$\pm$0.495 &9.15  &  & 146.18$\pm$    2.20    &42.25 $\pm$0.018          &40.97         \\
172.47 & 22.26 &0.291 & 7692.38$\pm$    1806.58 &42.99$\pm$0.318 &8.49  &  & 114.70$\pm$    9.59    &41.76 $\pm$0.192          &41.08         \\
246.01 & 11.34 &0.261 & 9794.67$\pm$     669.15 &43.38$\pm$0.936 &7.76  &  & 126.38$\pm$   17.38    &41.27 $\pm$0.142          &40.12         \\
126.91 &  9.70 &0.260 & 7538.43$\pm$     147.05 &43.06$\pm$0.075 &8.52  &  & 140.30$\pm$    3.42    &42.45 $\pm$0.033          &40.29         \\
150.74 & 16.04 &0.286 & 3783.67$\pm$     133.51 &43.60$\pm$0.590 &8.32  &  & 116.67$\pm$   61.90    &41.41 $\pm$0.570          &41.14         \\
202.55 & 18.44 &0.260 & 3847.71$\pm$      43.21 &43.50$\pm$0.133 &7.90  &  & 168.91$\pm$  153.55    &40.95 $\pm$1.301          &39.53         \\
192.84 & 21.42 &0.235 & 4743.97$\pm$      90.25 &43.10$\pm$0.214 &8.27  &  &  93.00$\pm$   87.28    &40.92 $\pm$1.246          &39.81         \\
186.41 & 24.98 &0.268 & 9059.90$\pm$     450.14 &43.16$\pm$0.180 &8.77  &  & 154.15$\pm$    1.75    &42.63 $\pm$0.015          &41.45         \\
194.53 & 23.49 &0.258 &10824.70$\pm$     138.97 &43.34$\pm$0.622 &8.28  &  & 311.54$\pm$  126.45    &40.23 $\pm$1.233          &39.97         \\
210.45 & 23.40 &0.215 & 5104.58$\pm$     130.12 &43.37$\pm$0.289 &8.86  &  & 324.15$\pm$  138.63    &41.30 $\pm$1.414          &41.00         \\
258.27 & 35.39 &0.084 & 2777.50$\pm$     290.11 &42.20$\pm$0.095 &7.22  &  & 212.83$\pm$   19.43    &40.75 $\pm$0.172          &39.41         \\
223.50 &  9.26 &0.279 & 9865.83$\pm$     630.65 &43.10$\pm$0.637 &7.75  &  &  91.08$\pm$    2.78    &41.88 $\pm$0.034          &41.59         \\
144.30 & 50.15 &0.275 & 1967.38$\pm$      59.63 &43.58$\pm$0.572 &7.65  &  & 137.81$\pm$   44.00    &41.35 $\pm$0.381          &41.69      \\ \hline
\hline
\end{tabular}  \\
\end{table*}

\end{document}